\documentclass{IEEEtran}
\usepackage{mathdots} 
\usepackage{amsmath,amsthm,amssymb,mathtools}
\usepackage{cite}
\usepackage[colorlinks=false]{hyperref}
\usepackage{graphicx}
\usepackage[caption=false,font=footnotesize,subrefformat=parens,labelformat=parens]{subfig}
\usepackage{threeparttable}

\usepackage{algorithmicx}
\usepackage{algorithm}
\usepackage[noend]{algpseudocode}

\usepackage[dvipsnames]{xcolor}
\usepackage{tikz}
\usetikzlibrary{shapes,arrows}
\graphicspath{{Figures/}}
\DeclareGraphicsExtensions{.eps}

\newtheorem{mylemma}{Lemma}

\newtheoremstyle{mytheoremstyle}{0pt}{0pt}{\itshape}{}{\bfseries}{.}{.5em}{} 
\theoremstyle{mytheoremstyle}

\makeatletter
\def\thmhead@plain#1#2#3{%
	\thmname{#1}\thmnumber{\@ifnotempty{#1}{ }\@upn{#2}}%
	\thmnote{ {\the\thm@notefont#3}}}
\let\thmhead\thmhead@plain
\def\thm@space@setup{\thm@preskip=0pt
	\thm@postskip=0pt}
\makeatother

\usepackage{xpatch}
\makeatletter
\xpatchcmd{\proof}{\@addpunct{.}}{\@addpunct{:}}{}{} 
\xpatchcmd{\proof}{\hskip\labelsep}{\hskip5\labelsep}{}{} 
\xpatchcmd{\proof}{\topsep6\p@\@plus6\p@\relax}{}{}{}
\makeatother

\usepackage{booktabs}

\allowdisplaybreaks

\begin{document}
\title{Effective Low-Complexity Optimization Methods for Joint Phase Noise and Channel Estimation in OFDM}
\author{Zhongju~Wang,
	Prabhu~Babu,
	and~Daniel~P.~Palomar,~\IEEEmembership{Fellow,~IEEE}
	\thanks{This work was supported by the Hong Kong RGC 16206315 research
		grant. Zhongju Wang and Daniel P. Palomar are with the Hong Kong University of Science and Technology (HKUST), Hong Kong. E-mail: {\{zwangaq, palomar\}}@ust.hk. Prabhu Babu is with CARE, IIT Delhi, Delhi, India. Email: prabhubabu@care.iitd.ac.in.}%
}


\maketitle

\begin{abstract}
	Phase noise correction is crucial to exploit full advantage of orthogonal frequency division multiplexing (OFDM) in modern high-data-rate communications. OFDM channel estimation with simultaneous phase noise compensation has therefore drawn much attention and stimulated continuing efforts. Existing methods, however, either have not taken into account the fundamental properties of phase noise or are only able to provide estimates of limited applicability owing to considerable computational complexity. In this paper, we have reformulated the joint estimation problem in the time domain as opposed to existing frequency-domain approaches, which enables us to develop much more efficient algorithms using the majorization-minimization technique. In addition, we propose a method based on dimensionality reduction and the Bayesian Information Criterion (BIC) that can adapt to various phase noise levels and accomplish much lower mean squared error than the benchmarks without incurring much additional computational cost. Several numerical examples with phase noise generated by free-running oscillators or phase-locked loops demonstrate that our proposed algorithms outperform existing methods with respect to both computational efficiency and mean squared error within a large range of signal-to-noise ratios.
\end{abstract}

%
\IEEEpeerreviewmaketitle
\section{Introduction}
\label{sec:Intro}
\IEEEPARstart{P}{ROMINENT} advantages such as higher spectral efficiency, adaptability to severe channel environments, and efficient implementation have brought orthogonal frequency division multiplexing (OFDM) into wide applications in modern communications. To fully exploit these advantages in reality, we have to resolve some demanding issues---sensitivity to frequency synchronization errors, high peak-to-average power ratios, to name a few. In this paper, we will focus on the frequency synchronization issue stemming specifically from phase noise.

Phase noise is a random process caused by the fluctuation within receiver and transmitter oscillators that are deployed to generate carrier signal for up-down conversion \cite{pollet1995ber,muschallik1995influence,tomba1998effect,lee2000oscillator,armada2001understanding,piazzo2002analysis,nikitopoulos2005phase}. In practice, free-running oscillators and phase-locked loops are widely used, for which phase noise is described by Wiener process and Gaussian process, respectively \cite{petrovic2007effects}. An OFDM block, consisting of several symbols, is transmitted and received with orthogonal subcarriers. Due to the interference of phase noise, however, the orthogonality among subcarriers is lost, which causes a degraded performance in OFDM systems. Indeed, common phase error (CPE) and inter-carrier interference (ICI) are two detrimental effects of phase noise. CPE causes subcarrier phase rotation that does not change within a transmitted OFDM block. In contrast, ICI introduces different interference to different subcarriers in the same block, and thus exhibits noise-like characteristics \cite{wu2004ofdm}. Generally, a constant carrier frequency offset (CFO) also exists apart from phase noise. With many methods available for CFO correction, herein we consider only phase noise estimation assuming CFO has been fixed; see, e.g., \cite{moose1994technique}.

Many works have studied phase noise estimation given known channel information, which is impractical because channel needs to be estimated as well. Subsequently, joint estimation of phase noise and channel impulse is proposed; see \cite{lin2006joint,zou2007compensation} and references therein. Such joint estimation problem has been investigated as early as in \cite{wu2003ofdm}. The least-squares estimator of channel impulse response is computed first; then heuristically, a window function as a filter is applied to the obtained channel estimator to reduce its sensitivity from phase noise and CFO. To be statistically justified, maximum \emph{a posteriori} channel estimator in \cite{lin2006joint} has exploited the statistical properties of phase noise. But the authors use Taylor expansion to approximate the nonlinear optimization objective function, which works only for small phase noise. A simple alternating optimization method for the joint estimation problem can be found in \cite{zou2007compensation}. The critical issue with that method is its failure to deal with the constraint of phase noise in each iterative sub-problem. Supposedly, estimating phase noise and channel was hard to disentangle as previous works claimed. In \cite{rabiei2010non}, an elegant formulation is proposed with phase noise and channel estimations unraveled. The authors replace the unimodular constraint on phase noise in the time domain with a relaxation assuming the magnitude of phase noise is relatively small. Nevertheless, their method is computationally unstable with a singularity issue that renders the already approximated solution even more inaccurate. And recently, a method craftily using the spectral property of phase noise is provided for the frequency domain-formulated problem \cite{mathecken2016ofdm}. Based on \cite{rabiei2010non}, the separate phase noise estimation problem is solved by semidefinite programming (SDP). This method woks fine when the number of subcarriers deployed in OFDM is not too large and phase noise arise in a small level. In reality, however, the number of subcarriers can be as large as tens of thousands, e.g., in terrestrial television broadcasting system (DVB-T2) \cite{etsi2009302}.

Regarding the joint phase noise and channel estimation, there are basically two ways for this problem: time-domain \cite{lin2006joint,zou2007compensation,wu2003ofdm,casas2002time,septier2007ofdm} and frequency-domain approaches \cite{petrovic2007effects,rabiei2010non,mathecken2016ofdm}. In this paper, we formulate the optimization problem in the time-domain representation and our contributions are as follows. First, we prove the equivalence of the frequency-domain approach and the time-domain approach to the problem formulation. It allows us to separate the joint estimation problem and to be focused on estimating phase noise. And using the majorization-minimization technique, we devise more efficient algorithms as opposed to solving an SDP as in \cite{mathecken2016ofdm}. The efficiency and low-complexity of our proposed algorithms enable us to deal readily with much larger number of subcarriers. Moreover, we offer an adaptive method for phase noise estimation considering that the joint estimation problem is underdetermined per se irrespective of the approach of formulation. To achieve this, dimensionality reduction has been adopted in \cite{zou2007compensation,mathecken2016ofdm} to address either the underdetermined nature of the problem or the computational complexity. But the algorithms provided therein cannot be easily extended to find the optimal reduced dimension. Instead of adhering to a presumed dimension, we implement our developed algorithms combined with Bayesian Information Criterion (BIC) and opt for the solution that yields the minimal BIC. This extra adaptability to various level of phase noise comes without incurring much computational burden as simulated examples demonstrate.

The structure of this paper is as follows. We give the system model of OFDM with a description of phase noise in Section \ref{sec:System_Model_Phase_Noise}. In Section \ref{sec:Lit_PNE}, the problem formulation is presented after a review of existing methods. We dedicate Section \ref{sec:Algorithms} to developing algorithms to solve the formulated problem with dimensionality reduction and BIC. Simulation results are given in Section \ref{sec:simulaitons}, followed by a conclusion to summarize the paper in Section \ref{sec:Conclusion}.

We use the following notation throughout this paper. Scalars, vectors, and matrices are denoted by italic letters, boldface lower-case letters, and boldface upper-case letters, respectively. The superscript $(\cdot)^T$ denotes the transpose, $(\cdot)^H$ the conjugate transpose. The $\ell_{2}$-norm and $\ell_{\infty}$-norm of a vector is denoted by $\|\cdot\|$ and $\|\cdot\|_{\infty}$, respectively. The identity matrix is denoted by $\mathbf{I}_{n}$ with size specified by the subscript $n$. $\mathbb{R}$ is the set of real numbers. $\mathbf{1}_n$ is an all-ones vector of length $n$. $\lambda_{\mathrm{max}}(\cdot)$ denotes the maximum eigenvalue of a matrix.

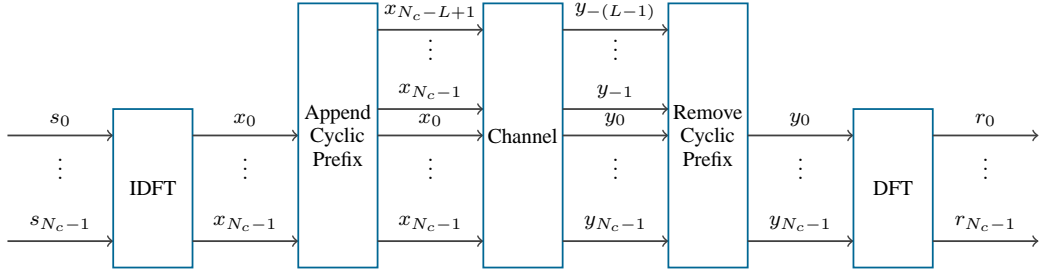
\begin{figure*}[!t]
	\centering
	\begin{tikzpicture}[line width=0.15ex]
	\tikzstyle{every node}=[font=\footnotesize]
	
	\draw [->, darkgray] (1em, 1em) -- (5em, 1em);
	\node[above] at (3em, 1em) {$s_{N_c-1}$};
	\draw [->, darkgray] (1em, 5em) -- (5em, 5em);
	\node[above] at (3em, 5em) {$s_{0}$};
	\node[below] at (3em, 5em) {$\vdots$};
	\draw [MidnightBlue] (5em, 0em) rectangle (8em, 6em);
	\node at (6.5em, 3em) {IDFT};
	\draw [->, darkgray] (8em, 1em) -- (12em, 1em);
	\node[above] at (10em, 1em) {$x_{N_c-1}$};
	\draw[->, darkgray] (8em, 5em) -- (12em, 5em);
	\node[above] at (10em, 5em) {$x_{0}$};
	\node[below] at (10em, 5em) {$\vdots$};
	\draw [MidnightBlue] (12em, 0em) rectangle (15em, 10em);
	\node[align=center] at (13.5em, 5em) {Append\\Cyclic\\Prefix};
	\draw [->, darkgray] (15em, 1em) -- (19em, 1em);
	\node[above] at (17em, 1em) {$x_{N_c-1}$};
	\draw[->, darkgray] (15em, 5em) -- (19em, 5em);
	\node[below] at (17em, 5em) {$\vdots$};
	\node [above] at (17em, 5em) {$x_{0}$};
	\draw[->, darkgray] (15em, 6em) -- (19em, 6em);
	\node[above] at (17em, 6em) {$x_{N_c-1}$};
	\draw[->, darkgray] (15em, 9em) -- (19em, 9em);
	\node[above] at (17em, 9em) {$x_{N_c-L+1}$};
	\node[below] at (17em, 9.5em) {$\vdots$};
	\draw [MidnightBlue] (19em, 0em) rectangle (22em, 10em);
	\node[align=center] at (20.5em, 5em) {Channel};
	\draw [->, darkgray] (22em, 1em) -- (26em, 1em);
	\node[above] at (24em, 1em) {$y_{N_c-1}$};
	\draw[->, darkgray] (22em, 5em) -- (26em, 5em);
	\node[below] at (24em, 5em) {$\vdots$};
	\node[above] at (24em, 5em) {$y_0$};
	\draw[->, darkgray] (22em, 6em) -- (26em,6em);
	\node[above] at (24em, 6em) {$y_{-1}$};
	\draw[->, darkgray] (22em, 9em) -- (26em, 9em);
	\node[below] at (24em, 9.5em) {$\vdots$};
	\node[above] at (24em, 9em) {$y_{-(L-1)}$};
	\draw [MidnightBlue] (26em, 0em) rectangle (29em, 10em);
	\node[align=center] at (27.5em, 5em) {Remove\\Cyclic\\Prefix};
	\draw [->, darkgray] (29em, 1em) -- (33em, 1em);
	\node[above] at (31em, 1em) {$y_{N_c-1}$};
	\draw[->, darkgray] (29em, 5em) -- (33em, 5em);
	\node[below] at (31em, 5em) {$\vdots$};
	\node[above] at (31em, 5em) {$y_0$};
	\draw[MidnightBlue] (33em, 0em) rectangle (36em, 6em);
	\node[align=center] at (34.5em, 3em) {DFT};
	\draw[->, darkgray] (36em, 1em) -- (40em, 1em);
	\node[above] at (38em, 1em) {$r_{N_c-1}$};
	\draw[->, darkgray] (36em, 5em) -- (40em, 5em);
	\node[below] at (38em, 5em) {$\vdots$};
	\node[above] at (38em, 5em) {$r_0$};
	\end{tikzpicture}
	\caption{Illustration of transmission and reception of OFDM.}
	\label{OFDM_trans_recp}
\end{figure*}

\section{System Model and Description of Phase Noise}
\label{sec:System_Model_Phase_Noise}

\subsection{OFDM Transmission Model}
\label{ssec:OFDM_model}
Suppose there are $N_c$ subcarriers and an OFDM block is denoted by $\mathbf{s}=\left[s_0, \dots,s_{N_c-1}\right]^T$. The time-domain symbols can be obtained by the unitary inverse discrete Fourier transform (IDFT):
\begin{equation}
\label{eq:IDFT_def}
x_n = \frac{1}{\sqrt{N_c}}\sum_{k=0}^{N_c-1}s_ke^{\frac{j2\pi nk}{N_c}},\qquad n=0,1,\dots,N_c-1.
\end{equation}
Let $\mathbf{F}$ be the $N_c\times N_c$ unitary discrete Fourier transform (DFT) matrix, then \eqref{eq:IDFT_def} can be written as
\begin{equation}
\label{eq:IDFT_def_mat}
\mathbf{x} = \mathbf{F}^H\mathbf{s}.
\end{equation}

Suppose the channel is linear time-invariant of length $L$ $(N_c\gg L)$, denoted by $\mathbf{h}=\left[h_0,h_1,\dots,h_{L-1}\right]^T$. To overcome the inter-symbol interference (ISI), a cyclic prefix of length at least $L-1$ is appended to the time-domain OFDM block $\mathbf{x}$ to be transmitted through the channel. We choose the minimum required length of cyclic prefix, i.e., $L-1$, and the actual transmitted symbols are $\left\{x_{N_c-L+1},\dots,x_{N_c-1},x_0,\dots,x_{N_c-1}\right\}$. The received symbols are $\left\{y_{-(L-1)},\dots,y_{-1},y_0,\dots,y_{N_c-1}\right\}$ and the first $L-1$ symbols, contaminated by the previous block, are discarded. Let $\mathbf{x}=\left[x_0,x_1,\dots,x_{N_c-1}\right]^T$ and $\mathbf{y}=\left[y_0,y_1,\dots,y_{N_c-1}\right]^T$. With the cyclic prefix appending and removal, we have the OFDM transmission model \cite[Ch. 3.4.4]{tse2005fundamentals}:
\begin{equation}
\label{eq:OFDM_time_domain}
\mathbf{y} = \mathbf{x}\circledast  \begin{bmatrix}
	\mathbf{h}\\\mathbf{0}
	\end{bmatrix} + \mathbf{v},
\end{equation}
where $\circledast $ denotes the operation of cyclic convolution, and $\mathbf{v}=\left[v_0,v_1,\dots,v_{N_c-1}\right]^T$ is a zero-mean circularly symmetric complex Gaussian channel noise vector with distribution $\mathcal{CN}(\mathbf{0},2\sigma^2\mathbf{I})$.

To obtain the frequency-domain representation of \eqref{eq:OFDM_time_domain}, take the DFT to both sides and we have\footnotemark[1]
\begin{equation}
\label{eq:OFDM_freq_domain}
\mathbf{r} = \sqrt{N_c}\mathbf{H}\mathbf{s} + \mathbf{w},
\end{equation}
where $\mathbf{r}$ is the unitary DFT of the received time-domain symbols $\mathbf{y}$, $\mathbf{H}$ is a diagonal matrix with the $N_c$-point unitary DFT of $\mathbf{h}$ as the diagonal, and $\mathbf{w}$ is the unitary DFT of the time-domain channel noise $\mathbf{v}$. Let $\check{\mathbf{F}}$ be a semi-unitary matrix formed by the first $L$ columns of $\mathbf{F}$, then $\mathbf{H}=\mathrm{Diag}\left(\check{\mathbf{F}}\mathbf{h}\right)$. The transmission and reception of OFDM are illustrated in Fig. \ref{OFDM_trans_recp}.
\footnotetext[1]{Note that the factor $\sqrt{N_c}$ results from using the unitary DFT.}

\subsection{OFDM Transmission with Phase Noise}
\label{ssec:Phase_Noise_in_OFDM}
In general, phase noise is present in the local oscillators that generate carrier signals for up-down conversion for the time-domain symbols. And the effect of phase noise can be represented mathematically by multiplying each time-domain symbol with a complex exponential with a random phase. Although phase noise exists in both the transmitter and the receiver, herein only the effect at the receiver side is studied. The reason for this simplified consideration is the assumption that at the transmitter side, the bandwidth of phase noise is small \cite{wu2004ofdm} or high-caliber oscillators are employed \cite{mathecken2016ofdm}. Therefore, the following signal model with phase noise is considered \cite{petrovic2007effects}:
\begin{equation}
\label{eq:OFDM_time_domain_PN}
\mathbf{y} = e^{j\boldsymbol{\theta}}\odot (\mathbf{x}\circledast \begin{bmatrix}
\mathbf{h}\\\mathbf{0}
\end{bmatrix}) + \mathbf{v},
\end{equation}
where $e^{j\boldsymbol{\theta}}\coloneqq\left[e^{j\theta_0},e^{j\theta_1},\dots,e^{j\theta_{N_c-1}}\right]^T$ denotes phase noise, and $\odot$ denotes the Hadamard product. Taking the unitary DFT on both sides of \eqref{eq:OFDM_time_domain_PN}, we can obtain the frequency-domain signal model:
\begin{equation}
\label{eq:OFDM_freq_domain_PN}
\mathbf{r} = \boldsymbol{\phi}\circledast  \left(\mathbf{H}\mathbf{s}\right) + \mathbf{w},
\end{equation}
where $\boldsymbol{\phi}=\left[\phi_0,\phi_1,\dots,\phi_{N_c-1}\right]^T=\mathbf{F}e^{j\boldsymbol{\theta}}$, called spectral phase noise vector, and $\mathbf{w}$ are the unitary DFT of $e^{j\boldsymbol{\theta}}$ and $\mathbf{v}$, respectively. For each received frequency-domain symbol $r_k, k=0,1,\dots,N_c-1$, we have
\begin{equation}
\label{eq:OFDM_freq_domain_PN_1}
r_k = \phi_{0}H_{k,k}s_k + \sum_{l=0,l\neq k}^{N_c-1}\phi_{k-l}H_{l,l}s_{l} + w_k,
\end{equation}
where the first term, subjected only to the scaling of factor $\phi_0$, is called CPE, and the second term, combining effects from other subcarriers, is ICI. With $r_k = r_{k\mod N_c}$, \eqref{eq:OFDM_freq_domain_PN} can be rewritten in the following matrix form
\begin{equation}
\label{eq:OFDM_freq_domain_PN_matrix_1}
\mathbf{r} = \boldsymbol{\Phi}_{\boldsymbol{\phi}}\mathbf{Hs} + \mathbf{w},
\end{equation}
in which
\begin{align}
\boldsymbol{\Phi}_{\boldsymbol{\phi}} & = \begin{bmatrix}
\phi_0 & \phi_{N_c-1} & \cdots & \phi_{2} & \phi_{1}\\
\phi_1 & \phi_0 & \cdots & \phi_{3} & \phi_{2}\\
\vdots & & \ddots & & \vdots\\
\phi_{N_c-2} & \phi_{N_c-3} & \cdots & \phi_{0} & \phi_{N_c-1}\\
\phi_{N_c-1} & \phi_{N_c-2} & \cdots & \phi_{1} & \phi_{0}
\end{bmatrix}\label{eq:PN_freq_circ_matrix_1},
\end{align}
denoted by $\boldsymbol{\Phi}_{\boldsymbol{\phi}}=\mathrm{circ}\left(\boldsymbol{\phi}\right)$, is a circulant matrix formed by spectral phase noise $\boldsymbol{\phi}$. The off-diagonals of $\boldsymbol{\Phi}_{\boldsymbol{\phi}}$ close to the main diagonal correspond to low-frequency components. With $\mathbf{H}=\mathrm{Diag}\left(\check{\mathbf{F}}\mathbf{h}\right)$, \eqref{eq:OFDM_freq_domain_PN_matrix_1} can be rewritten as
\begin{equation}
\label{eq:OFDM_freq_model_h}
\mathbf{r} = \boldsymbol{\Phi}_{\boldsymbol{\phi}}\mathbf{S}\check{\mathbf{F}}\mathbf{h} + \mathbf{w},
\end{equation}
where $\mathbf{S}=\mathrm{Diag}(\mathbf{s})$ is a diagonal matrix with $\mathbf{s}$ as the diagonal.

\subsection{Properties of Phase Noise}
\label{ssec:Properties_PN}
Two canonical models of phase noise are Wiener process and Gaussian process when free-running oscillators and phase-locked loops are respectively employed \cite{petrovic2007effects}. The statistical properties of phase noise have also been studied in \cite{wu2004ofdm,petrovic2007effects}. Before introducing some existing formulations of the phase noise estimation problem, some useful properties of phase noise are presented here.

\subsubsection{Time-Domain Property}
Obviously, phase noise $e^{j\boldsymbol{\theta}}$ is determined only by the phase variable $\boldsymbol{\theta}$, and phase noise at each OFDM subcarrier is {\em unimodular}, i.e.,
\begin{equation}
\label{eq:Time_property_PN}
\left|e^{j\theta_n}\right| = 1,\qquad~n=0,1,\dots,N_c-1.
\end{equation}

\subsubsection{Frequency-Domain Property}
Let $\boldsymbol{\phi}$ and $\underline{\boldsymbol{\phi}}$ be the unitary DFT of $e^{j\boldsymbol{\theta}}$ and $e^{-j\boldsymbol{\theta}}$, respectively. It is well-known that $\boldsymbol{\phi}$ and $\underline{\boldsymbol{\phi}}$ are conjugate symmetric, i.e.,
\begin{equation}
\label{eq:complex_exp_conj_sym}
\underline{\phi}_k=\phi_{-k}^\ast. 
\end{equation}
Observing that $e^{j\boldsymbol{\theta}}\odot e^{-j\boldsymbol{\theta}} = \mathbf{1}_{N_c}$ and applying the DFT to both sides, we can obtain the following constraint for spectral phase noise:
\begin{equation}
\label{eq:Freq_property_PN}
\boldsymbol{\phi}\circledast  \underline{\boldsymbol{\phi}} = N_c\delta_k,
\end{equation}
where $\delta_k$ is the Kronecker delta function, i.e., $\delta_0 = 1$, and $\delta_k=0$ for $k\neq 0$. Indeed, \eqref{eq:Freq_property_PN} is a necessary and sufficient description of the autocorrelation of the spectral components of any unimodular complex exponential sequence, which can be easily verified by Fourier transform and its properties. Equivalently, \eqref{eq:Freq_property_PN} can be written in a matrix form as
\begin{equation}
\label{eq:Freq_property_PN_matrix}
\boldsymbol{\Phi}_{\boldsymbol{\phi}}^H\boldsymbol{\Phi}_{\boldsymbol{\phi}} = N_c\mathbf{I}_{N_c},
\end{equation}
where $\boldsymbol{\Phi}_{\boldsymbol{\phi}}=\mathrm{circ}\left(\boldsymbol{\phi}\right)$ is a circulant matrix defined in \eqref{eq:PN_freq_circ_matrix_1}. This is the main property exploited in \cite{mathecken2016ofdm}, termed the spectral geometry.

\section{Literature Review and Problem Formulation}
\label{sec:Lit_PNE}

Phase noise contamination can be removed from the received OFDM symbols if a reliable estimate of the instantaneous realization of phase noise process is accessible. But a thorny issue is that phase noise estimation is entangled with the unknown channel and even further the unknown transmitted data. Many works have studied both scenarios, but in this paper we will be focused on joint phase noise and channel estimation. A motivation is that assuming channel is quasistatic or slowly-varying, a channel estimate can be used in the subsequent data detection. For the joint estimation problem, methods in the literature can be categorized into two classes: time-domain approach and frequency-domain approach. Throughout the paper, we assume phase noise $\boldsymbol{\theta}$ and channel impulse response $\mathbf{h}$ are constant parameters, and OFDM symbols $\mathbf{S}$ are given and known to the receiver.

\subsection{Time-Domain Approaches}
In \cite{zou2007compensation}, the authors formulate the least-squares problem with \eqref{eq:OFDM_freq_model_h}
\begin{equation}
	\label{eq:TD_PN_LS_alt_MM_prob}
	\begin{array}{rl}
		\underset{\mathbf{h},\boldsymbol{\theta},\boldsymbol{\phi}=\mathbf{F}e^{j\boldsymbol{\theta}}}{\text{minimize}} & \left\|\mathbf{r}-\boldsymbol{\Phi}_{\boldsymbol{\phi}}\mathbf{S}\check{\mathbf{F}}\mathbf{h}\right\|^2,
	\end{array}
\end{equation}
and solve for channel and phase noise estimates alternately. At the $i$th iteration, given the phase noise estimate $e^{j\hat{\boldsymbol{\theta}}^{(i-1)}}$, the channel estimate is computed by
\begin{equation}
\label{eq:least_squares_h}
	\hat{\mathbf{h}}^{(i)} = \left(\check{\mathbf{F}}^{H}\mathbf{S}^{H}(\boldsymbol{\Phi}_{\boldsymbol{\phi}}^{(i-1)})^H\boldsymbol{\Phi}_{\boldsymbol{\phi}}^{(i-1)}\mathbf{S}\check{\mathbf{F}}\right)^{-1}\check{\mathbf{F}}^{H}\mathbf{S}^{H}(\boldsymbol{\Phi}_{\boldsymbol{\phi}}^{(i-1)})^{H}\mathbf{r}
\end{equation}
with $\boldsymbol{\Phi}_{\boldsymbol{\phi}}^{(i-1)} = \mathrm{circ}(\mathbf{F}e^{j\hat{\boldsymbol{\theta}}^{(i-1)}})$. Let $\mathbf{c} = e^{j\mathbf{\boldsymbol{\theta}}}$, then the estimate for phase noise is updated as
\begin{equation}
\label{eq:least_squares_c}
	\hat{\mathbf{c}}^{(i)}  = \left(\mathbf{F}^H\mathbf{P}^H\mathbf{P}\mathbf{F}\right)^{-1}\mathbf{F}^H\mathbf{P}^H\mathbf{r},
\end{equation}
where $\mathbf{P}=\mathrm{circ}(\mathbf{S}\check{\mathbf{F}}\hat{\mathbf{h}}^{(i)})$. Yet there are two issues with their method: the unimodulus property of phase noise vector is not considered when updating $\hat{\mathbf{c}}^{(i)}$; and the alternating optimization scheme suffers from slow convergence. 

Some other heuristic methods include approximating phase noise by Taylor expansions \cite{lin2006joint}, applying filtering to channel estimate with a noise-suppressing function \cite{wu2003ofdm}, approximating with sinusoidal waveforms \cite{casas2002time}, and Monte Carlo methods \cite{septier2007ofdm}.

\subsection{Frequency-Domain Approaches}
\label{ssub:PN_correction_spectral}
In \cite{petrovic2007effects}, a phase noise correction method is proposed by estimating the spectral components, based on the assumption that phase noise process can be characterized by a low-pass signal and thus only a few spectral components need to be estimated. But it is necessary to find a proper number of spectral phase noise components in order to achieve reliable estimation. Although \cite{petrovic2007effects} also exploits the statistical properties of ICI to obtain the MMSE estimate of phase noise, their method is subject to two main issues: the channel is assumed known, and the MMSE estimation has not taken into account the constraint \eqref{eq:Freq_property_PN} of spectral phase noise.

Following the same idea of \cite{petrovic2007effects} to estimate the low-frequency components of phase noise, \cite{rabiei2010non} formulates the problem of joint phase noise and channel estimation based on least-squares. To acquire separate estimators, instead of alternately updating \eqref{eq:least_squares_h} and \eqref{eq:least_squares_c}, they substitute the channel estimate into the least-squares objective and the resulting error function for phase noise can be derived as
\begin{align}
	\mathcal{E}\left(\boldsymbol{\phi}\right) & = \mathbf{r}^{H}\mathbf{r} - \frac{1}{N_c}\mathbf{r}^{H}\boldsymbol{\Phi}_{\boldsymbol{\phi}}\mathbf{B}\boldsymbol{\Phi}_{\boldsymbol{\phi}}^{H}\mathbf{r}\label{eq:least_squares_err_Phi_2}\\
	& = \frac{1}{N_c}\boldsymbol{\phi}^{H}\mathbf{J}_1\left(\mathbf{R}^{H}\mathbf{R} - \mathbf{R}^{H}\mathbf{B}\mathbf{R}\right)^{T}\mathbf{J}_1\boldsymbol{\phi},\label{eq:least_squares_err_Phi_8}
\end{align}
where $\mathbf{B} = \mathbf{S}\check{\mathbf{F}}\left(\check{\mathbf{F}}^{H}\mathbf{S}^{H}\mathbf{S}\check{\mathbf{F}}\right)^{-1}\check{\mathbf{F}}^{H}\mathbf{S}^{H}$, and $\mathbf{J}_1$ a permutation matrix defined as
\begin{equation}
\mathbf{J}_1 = \begin{bmatrix}
1 & & & & \\
& & & & 1\\
& & & 1 &\\
& & \iddots & &\\
& 1 & & &
\end{bmatrix}.
\end{equation}
Note that in \cite{rabiei2010non}, the expression for $\mathcal{E}\left(\boldsymbol{\phi}\right)$ is further simplified assuming the transmitted symbols are of constant-modulus. When solving for the phase noise estimate, however, an approximation by Taylor expansion is applied, which leads to a relaxed constraint on phase noise. In practice, this approximation works only for small phase noise. 

In contrast, \cite{mathecken2016ofdm} incorporates the fundamental spectral constraint \eqref{eq:Freq_property_PN_matrix} into the formulation proposed in \cite{rabiei2010non}. Let
\begin{equation}
\mathbf{M} = \frac{1}{N_c}\mathbf{J}_1\left(\mathbf{R}^{H}\mathbf{R} - \mathbf{R}^{H}\mathbf{B}\mathbf{R}\right)^{T}\mathbf{J}_1,
\end{equation}
then the problem is formulated as
\begin{equation}
\label{eq:min_LS_err_spectral}
\begin{array}{rl}
\underset{\boldsymbol{\phi}}{\text{minimize}} & \boldsymbol{\phi}^{H}\mathbf{M}\boldsymbol{\phi}\\
\text{subject to} & \boldsymbol{\Phi}_{\boldsymbol{\phi}}^H\boldsymbol{\Phi}_{\boldsymbol{\phi}}=N_c\mathbf{I}_{N_c}, \boldsymbol{\Phi}_{\boldsymbol{\phi}}=\mathrm{circ}\left(\boldsymbol{\phi}\right).
\end{array}
\end{equation}
Instead of solving \eqref{eq:min_LS_err_spectral}, dimensionality reduction is introduced to alleviate the computation complexity by estimating only the low-frequency components, cf. \cite{petrovic2007effects}. To achieve this, the phase-noise-geometry preserving transformation is defined by
\begin{equation}
\label{eq:transformation_spectral}
	\boldsymbol{\phi} = \mathbf{T}\check{\boldsymbol{\phi}},
\end{equation}
where $\check{\boldsymbol{\phi}}$ of a shorter length $N$ is the reduced spectral phase noise to be estimated. An example of $\mathbf{T}$ is piecewise-constant transformation (PCT). Then an alternative optimization problem is posed as follows:
\begin{equation}
\label{eq:min_LS_err_spectral_reduced}
\begin{array}{rl}
	\underset{\check{\boldsymbol{\phi}}}{\text{minimize}} & \check{\boldsymbol{\phi}}^{H}\mathbf{T}^{H}\mathbf{M}\mathbf{T}\check{\boldsymbol{\phi}}\\
	\text{subject to} & \check{\boldsymbol{\Phi}}^H_{\boldsymbol{\phi}}\check{\boldsymbol{\Phi}}_{\boldsymbol{\phi}}=N\mathbf{I}_N,\check{\boldsymbol{\Phi}}_{\boldsymbol{\phi}}=\mathrm{circ}(\check{\boldsymbol{\phi}}).
\end{array}
\end{equation}
To solve the above problem, the S-procedure is invoked to rewrite \eqref{eq:min_LS_err_spectral_reduced} as a semidefinite program (SDP). The original spectral phase noise vector $\boldsymbol{\phi}$ can be recovered by \eqref{eq:transformation_spectral}. To guarantee the constraint \eqref{eq:Freq_property_PN_matrix} still holds, the authors provide a sufficient condition for the transformation matrix. Their method, however, suffers from several limitations. When the reduced length $N$ is not small enough, SDP reformulation still renders a solution failing to satisfy the spectral constraint of phase noise; yet, it is prohibited to solve a large dimensional SDP. Nowadays, the number of subcarriers can be up to thousands and to use this method, the original dimension needs to be greatly reduced, which can result in the loss of reliability and accuracy in the obtained estimate. Furthermore, the reduced spectral phase noise does not necessarily satisfy the spectral constraint as imposed in problem \eqref{eq:min_LS_err_spectral_reduced}; thus, this method gives a tightened solution.

\subsection{Problem Formulation}
\label{ssec:Problem_Formulation}
Based on \eqref{eq:OFDM_time_domain_PN}, the time-domain OFDM model with phase noise is given by
\begin{equation}
\label{eq:time_domain_OFDM_PN}
	\mathbf{y} = \sqrt{N_c}\mathrm{Diag}\left(e^{j\boldsymbol{\theta}}\right)\mathbf{F}^{H}\mathbf{S}\check{\mathbf{F}}\mathbf{h} + \mathbf{v}.
\end{equation}
Similar to \eqref{eq:TD_PN_LS_alt_MM_prob}, we propose the following optimization problem:
\begin{equation}
\label{eq:TD_PN_LS_prob}
	\begin{array}{rl}
	\underset{\mathbf{h},\boldsymbol{\theta}}{\text{minimize}} & \left\|\mathbf{y}-\sqrt{N_c}\mathrm{Diag}\left(e^{j\boldsymbol{\theta}}\right)\mathbf{F}^{H}\mathbf{S}\check{\mathbf{F}}\mathbf{h}\right\|^2.
	\end{array}
\end{equation}
Solving \eqref{eq:TD_PN_LS_prob} for $\mathbf{h}$ gives the least-squares channel estimate
\begin{equation}
	\label{eq:TD_PN_LS_CIR}
	\hat{\mathbf{h}} = \frac{1}{\sqrt{N_c}}\left(\check{\mathbf{F}}^{H}\mathbf{S}^H\mathbf{S}\check{\mathbf{F}}\right)^{-1}\check{\mathbf{F}}^{H}\mathbf{S}^H\mathbf{F}\mathrm{Diag}\left(e^{j\boldsymbol{\theta}}\right)^H\mathbf{y}.
\end{equation}
And the resulting least-squares error for phase noise is
\begin{equation}
\label{eq:LS_error_theta}
	\mathcal{E}(\boldsymbol{\theta}) = \mathbf{y}^H\mathrm{Diag}\left(e^{j\boldsymbol{\theta}}\right)\mathbf{F}^{H}\left(\mathbf{I}_{N_c} - \mathbf{B}\right)\mathbf{F}\mathrm{Diag}\left(e^{j\boldsymbol{\theta}}\right)^H\mathbf{y},
\end{equation}
where $\mathbf{B} = \mathbf{S}\check{\mathbf{F}}\left(\check{\mathbf{F}}^{H}\mathbf{S}^{H}\mathbf{S}\check{\mathbf{F}}\right)^{-1}\check{\mathbf{F}}^{H}\mathbf{S}^{H}$. The phase noise estimation problem is thus formulated as
\begin{equation}
\label{eq:min_LS_err_time_domain_}
	\begin{array}{rl}
		\underset{\boldsymbol{\theta}}{\text{minimize}} & \mathbf{y}^H\mathrm{Diag}\left(e^{j\boldsymbol{\theta}}\right)\mathbf{F}^{H}\left(\mathbf{I}_{N_c} - \mathbf{B}\right)\mathbf{F}\mathrm{Diag}\left(e^{j\boldsymbol{\theta}}\right)^H\mathbf{y}.
	\end{array}
\end{equation}
Let us introduce $\mathbf{V}=\mathbf{F}^{H}\left(\mathbf{I}_{N_c} - \mathbf{B}\right)\mathbf{F}$ and $\mathbf{u} = e^{-j\boldsymbol{\theta}}$. We can rewrite \eqref{eq:min_LS_err_time_domain_} as the following quadratic problem:
\begin{equation}
\label{eq:min_LS_err_time_domain}
	\begin{array}{rl}
	\underset{\mathbf{u}}{\text{minimize}} & \mathbf{u}^H\mathrm{Diag}(\mathbf{y})^H\mathbf{V}\mathrm{Diag}(\mathbf{y})\mathbf{u}\\
	\text{subject to} & |u_n| = 1,\qquad n=0,1,\dots,N_c-1.
	\end{array}
\end{equation}
Consequently, the joint phase noise and channel estimation problem boils down to the phase noise estimation problem \eqref{eq:min_LS_err_time_domain} followed by computing the channel estimate with \eqref{eq:TD_PN_LS_CIR}.

\subsection{Equivalence of Time- and Frequency-Domain Approaches}
\label{ssec:equivalence_time_freq_approach}
In this section, we show that our formulation of the joint estimation problem \eqref{eq:TD_PN_LS_prob} and the resulting phase noise estimation problem \eqref{eq:min_LS_err_time_domain} are equivalent to the existing approaches.

\begin{mylemma}\label{lemma2}
	Let $\boldsymbol{\phi}=\mathbf{F}e^{j\boldsymbol{\theta}}$ and $\boldsymbol{\Phi}_{\boldsymbol{\phi}}=\mathrm{circ}(\boldsymbol{\phi})$, then $\boldsymbol{\Phi}_{\boldsymbol{\phi}}=\sqrt{N_c}\mathbf{F}\mathrm{Diag}\left(e^{j\boldsymbol{\theta}}\right)\mathbf{F}^H$.
	\begin{proof}
		According to the eigenvalue decomposition of a circulant matrix \cite{gray2006toeplitz},
		\begin{align}
		\boldsymbol{\Phi}_{\boldsymbol{\phi}} & = \mathbf{F}\mathrm{Diag}\left(\sqrt{N_c}\mathbf{F}\left(\mathbf{e}_1^T\boldsymbol{\Phi}_{\boldsymbol{\phi}}\right)^T\right)\mathbf{F}^H\nonumber\\
		& = \mathbf{F}\mathrm{Diag}\left(\sqrt{N_c}\mathbf{F}^H\left(\mathbf{e}_1^T\boldsymbol{\Phi}_{\boldsymbol{\phi}}\right)^H\right)^\ast\mathbf{F}^H\nonumber\\
		& = \mathbf{F}\mathrm{Diag}\left(\sqrt{N_c}\mathbf{F}^H\underline{\boldsymbol{\phi}}\right)^\ast\mathbf{F}^H\nonumber\\
		& = \mathbf{F}\mathrm{Diag}\left(\sqrt{N_c}e^{-j\boldsymbol{\theta}}\right)^\ast\mathbf{F}^H\nonumber\\
		& = \sqrt{N_c}\mathbf{F}\mathrm{Diag}\left(e^{j\boldsymbol{\theta}}\right)\mathbf{F}^H\nonumber,
		\end{align}
		where $\mathbf{e}_1=[1\ 0\ \cdots\ 0]^T$.
	\end{proof}
\end{mylemma}
With Lemma \ref{lemma2}, we can prove that the objective function in problem \eqref{eq:TD_PN_LS_prob} is the same as that of \eqref{eq:TD_PN_LS_alt_MM_prob}:
\begin{align}
	& \left\|\mathbf{y}-\sqrt{N_c}\mathrm{Diag}\left(e^{j\boldsymbol{\theta}}\right)\mathbf{F}^{H}\mathbf{S}\check{\mathbf{F}}\mathbf{h}\right\|^2\nonumber\\
	= & \left\|\mathbf{F}^H\mathbf{r}-\sqrt{N_c}\mathrm{Diag}\left(e^{j\boldsymbol{\theta}}\right)\mathbf{F}^{H}\mathbf{S}\check{\mathbf{F}}\mathbf{h}\right\|^2\\
	= & \left\|\mathbf{r}-\sqrt{N_c}\mathbf{F}\mathrm{Diag}\left(e^{j\boldsymbol{\theta}}\right)\mathbf{F}^{H}\mathbf{S}\check{\mathbf{F}}\mathbf{h}\right\|^2\\
	= & \left\|\mathbf{r}-\boldsymbol{\Phi}_{\boldsymbol{\phi}}\mathbf{S}\check{\mathbf{F}}\mathbf{h}\right\|^2.
\end{align}

Since
\begin{equation}
\label{eq:time_freq_LS_equality}
	\boldsymbol{\Phi}_{\boldsymbol{\phi}}^{H}\mathbf{r} = \sqrt{N_c}\mathbf{F}\mathrm{Diag}\left(e^{j\boldsymbol{\theta}}\right)^H\mathbf{F}^H\mathbf{r} = \sqrt{N_c}\mathbf{F}\mathrm{Diag}\left(e^{j\boldsymbol{\theta}}\right)^H\mathbf{y},
\end{equation}
\eqref{eq:LS_error_theta} is equivalent to the frequency-domain phase noise error function \eqref{eq:least_squares_err_Phi_2}:
\begin{align}
	\mathcal{E}\left(\boldsymbol{\phi}\right)
	& = \frac{1}{N_c}\mathbf{r}^{H}\boldsymbol{\Phi}_{\boldsymbol{\phi}}\left(\mathbf{I}_{N_c} - \mathbf{B}\right)\boldsymbol{\Phi}_{\boldsymbol{\phi}}^{H}\mathbf{r}\\
	& = \mathbf{y}^H\mathrm{Diag}\left(e^{j\boldsymbol{\theta}}\right)\mathbf{F}^{H}\left(\mathbf{I}_{N_c} - \mathbf{B}\right)\mathbf{F}\mathrm{Diag}\left(e^{j\boldsymbol{\theta}}\right)^H\mathbf{y}.
\end{align}
In the next section, we will use the majorization-minimization technique to develop efficient algorithms to solve problem \eqref{eq:min_LS_err_time_domain}.

\section{Algorithms}
\label{sec:Algorithms}
\subsection{The Majorization-Minimization Technique}
\label{ssec:Majorization_Minimization_Framework}
The majorization-minimization (MM) technique provides an approximation-based iterative approach to solving an optimization problem of a generic form \cite{hunter2004tutorial,beck2009gradient,sun2016mm}. As the original problem is difficult to address directly, the MM technique follows an iterative procedure---a simpler surrogate objective function is minimized in each iteration---to find a local optimum.

Consider the problem of
\begin{eqnarray}
\label{eq:MM_original_prob}
	\begin{array}{rcrc}
	\underset{\mathbf{x}}{\text{minimize}} & f(\mathbf{x}) & \text{subject to} & \mathbf{x}\in\mathcal{X}.
	\end{array}
\end{eqnarray}
The MM technique starts from a feasible point $\mathbf{x}^{(0)}\in\mathcal{X}$, and solves a series of simpler majorized problems:
\begin{eqnarray}
	\label{eq:MM_prob_t}
	\begin{array}{rcrc}
		\underset{\mathbf{x}}{\text{minimize}} & g\left(\mathbf{x};\mathbf{x}^{(t)}\right) & \text{subject to} & \mathbf{x}\in\mathcal{X},
	\end{array}
\end{eqnarray}
$t=0,1,\dots$, each of which produces an updated point $\mathbf{x}^{(t+1)}$. Basically, the surrogate objective, known as the majorization function for $f(\mathbf{x})$, should satisfy the following conditions:
\begin{align}
g\left(\mathbf{x}^{(t)};\mathbf{x}^{(t)}\right) & = f\left(\mathbf{x}^{(t)}\right),\label{eq:MM_majorizaiton_1}\\
g\left(\mathbf{x};\mathbf{x}^{(t)}\right) & \geq f\left(\mathbf{x}\right) \qquad\forall\mathbf{x}\in\mathcal{X},\label{eq:MM_majorizaiton_2}\\
\nabla_{\mathbf{d}}g\left(\mathbf{x}^{(t)};\mathbf{x}^{(t)}\right) & = \nabla_{\mathbf{d}}f\left(\mathbf{x}^{(t)}\right)\qquad\forall\mathbf{x}^{(t)}+\mathbf{d}\in\mathcal{X},
\end{align}
where $\nabla_{\mathbf{d}}g\left(\mathbf{x}^{(t)};\mathbf{x}^{(t)}\right)$ is the directional derivative of $g$ at $\mathbf{x}^{(t)}$ in the direction of $\mathbf{d}$. Consequently, a series of points that result in nonincreasing objective values are obtained:
\begin{equation}\label{eq:MM_inequs}
f\left(\mathbf{x}^{(t+1)}\right)\leq g\left(\mathbf{x}^{(t+1)};\mathbf{x}^{(t)}\right)\leq g\left(\mathbf{x}^{(t)};\mathbf{x}^{(t)}\right)=f\left(\mathbf{x}^{(t)}\right),
\end{equation}
And any limit point of thus generated sequence of points is a stationary solution to the original problem \eqref{eq:MM_original_prob}.

To develop an efficient MM-based algorithm, the series of problems \eqref{eq:MM_prob_t} should all be simple enough to solve---ideally, each can be solved with a closed-form solution. Crucial to achieve such a goal is to find a good majorization function $g\left(\mathbf{x};\mathbf{x}^{(t)}\right)$, which requires to properly exploit the particular structure of the specific problem. Some general and useful rules for majorization can be found in \cite{sun2016mm}. In the next section, we will devise MM algorithms to solve our problem \eqref{eq:min_LS_err_time_domain} with two different majorizing methods. Also it will be illustrated in simulations that the majorization is critical for the convergence speed of the obtained algorithms.

\subsection{The MM Algorithms for Phase Noise Estimation}
\label{ssec:MM_Algorithms_PN}
The following lemma is introduced first, which is useful for finding majorization functions.
\begin{mylemma}\label{lemma3}
	Given a matrix $\mathbf{A}$, $\mathbf{P} = \mathbf{A}(\mathbf{A}^H\mathbf{A})^{-1}\mathbf{A}^H$ is an orthogonal projection matrix, which is unitarily similar to a diagonal matrix with diagonal entries being either 1 or 0 \cite[Corollary 3.4.3.3]{horn2012matrix}.
\end{mylemma}

\subsubsection{Loose Quadratic Majorization (LQM)}
Let us write $\tilde{\mathbf{V}}=\mathrm{Diag}(\mathbf{y})^H\mathbf{V}\mathrm{Diag}(\mathbf{y})$. The objective in \eqref{eq:min_LS_err_time_domain} can be majorized by a quadratic function at $\mathbf{u}_0$ as follows \cite[Lemma 1]{song2014optimal}:
\begin{equation}
\label{eq:LQM_Majorization}
	\mathbf{u}^H\tilde{\mathbf{V}}\mathbf{u} \leq 2\mathrm{Re}\left\{\mathbf{u}_0^H\left(\tilde{\mathbf{V}}-\lambda\mathbf{I}_{N_c}\right)\mathbf{u}\right\} + 2\lambda\|\mathbf{u}\|^2 - \mathbf{u}_0^H\tilde{\mathbf{V}}\mathbf{u}_0,
\end{equation}
in which $\lambda\mathbf{I}_{N_c}\succeq\tilde{\mathbf{V}}$ for some constant $\lambda$. Note that the largest eigenvalue of $\mathbf{V}$ is 1 by Lemma \ref{lemma3}, then we have $\lambda_{\mathrm{max}}(\tilde{\mathbf{V}})\leq \|\mathbf{y}\|^2_{\infty}$. Choosing $\lambda=\|\mathbf{y}\|^2_{\infty}$ will thus satisfy the majorization condition. At the step $t$, the following majorized problem with the surrogate objective function is solved (since $\|\mathbf{u}\|^2$ is just a constant):
\begin{equation}
\label{eq:min_LS_err_time_domain_MM_2_prob}
	\begin{array}{rl}
	\underset{\mathbf{u}}{\text{minimize}} & -2\mathrm{Re}\left\{\left(\mathbf{u}^{(t)}\right)^H\left(\|\mathbf{y}\|^2_{\infty}\mathbf{I}_{N_c} - \tilde{\mathbf{V}}\right)\mathbf{u}\right\}\\
	\text{subject to} & |u_n| = 1,\qquad n=0,1,\dots,N_c-1.
	\end{array}
\end{equation}
It is obvious that a closed-form solution to \eqref{eq:min_LS_err_time_domain_MM_2_prob} is:
\begin{align}
	\mathbf{u}^{(t+1)} = &\,\exp\left[j\arg\left(\big(\|\mathbf{y}\|^2_{\infty}\mathbf{I}_{N_c} - \tilde{\mathbf{V}}\big)\mathbf{u}^{(t)}\right)\right]\\
	= &\,\exp\Big[j\arg\Big(\big(\|\mathbf{y}\|^2_{\infty}\mathbf{1}-|\mathbf{y}|^2\big)\odot\mathbf{u}^{(t)}\nonumber\\
	& + \mathrm{Diag}(\mathbf{y})^H\mathbf{F}^{H}\mathbf{B}\mathbf{F}\mathrm{Diag}(\mathbf{y})\mathbf{u}^{(t)}\Big)\Big]\label{eq:TD_MM_2_update},
\end{align}
where the exponential and the squared magnitude $|\cdot|^2$ are taken element-wise. We call this method a loose quadratic majorization (LQM) because the structure of the original objective function could have been better exploited as shown below, which leads to faster convergence.

\subsubsection{Tight Quadratic Majorization (TQM)}
Similar to \eqref{eq:LQM_Majorization}, the original objective can be majorized as follows:
\begin{align}
	& \mathbf{u}^H\mathrm{Diag}(\mathbf{y})^H\mathbf{V}\mathrm{Diag}(\mathbf{y})\mathbf{u} \nonumber\\[1pt]
	\leq\ & \lambda\mathbf{u}^H\mathrm{Diag}(\mathbf{y})^H\mathrm{Diag}(\mathbf{y})\mathbf{u} \nonumber\\[1pt]
	& + 2\mathrm{Re}\Big\{\mathbf{u}_0^H\mathrm{Diag}(\mathbf{y})^H\left(\mathbf{V}-\lambda\mathbf{I}_{N_c}\right)\mathrm{Diag}(\mathbf{y})\mathbf{u}\Big\} \nonumber\\[1pt]
	& + \mathbf{u}_0^H\mathrm{Diag}(\mathbf{y})^H\left(\lambda\mathbf{I}_{N_c}-\mathbf{V}\right)\mathrm{Diag}(\mathbf{y})\mathbf{u}_0\label{eq:min_LS_err_time_domain_MM_1}\\[1pt]
	=\ & 2\mathrm{Re}\Big\{\mathbf{u}_0^H\mathrm{Diag}(\mathbf{y})^H\left(\mathbf{V}-\lambda\mathbf{I}_{N_c}\right)\mathrm{Diag}(\mathbf{y})\mathbf{u}\Big\} \nonumber\\[1pt]
	& + 2\lambda \|\mathbf{y}\|^2 - \mathbf{u}_0^H\mathrm{Diag}(\mathbf{y})^H\mathbf{V}\mathrm{Diag}(\mathbf{y})\mathbf{u}_0,
\end{align}
where $\lambda\mathbf{I}_{N_c}\succeq\mathbf{V}$ for some constant $\lambda$ and the equality follows from the unimodulus of $u_n, n=0,1,\dots,N_c-1$. To find a good majorization function, we can choose $\lambda = 1$ by Lemma \ref{lemma3}. At the step $t$, the following majorized problem can be obtained:
\begin{equation}
\label{eq:min_LS_err_time_domain_MM_1_prob}
\begin{array}{rl}
\underset{\mathbf{u}}{\text{minimize}} & -2\mathrm{Re}\left\{\left(\mathbf{u}^{(t)}\right)^H\mathrm{Diag}(\mathbf{y})^H\mathbf{F}^{H}\mathbf{B}\mathbf{F}\mathrm{Diag}(\mathbf{y})\mathbf{u}\right\}\\
\text{subject to} & |u_n| = 1,\qquad n=0,1,\dots,N_c-1,
\end{array}
\end{equation}
which results in a closed-form solution:
\begin{equation}
\label{eq:TD_MM_1_update}
	\mathbf{u}^{(t+1)} = \exp\left[j\arg\left(\mathrm{Diag}(\mathbf{y})^H\mathbf{F}^{H}\mathbf{B}\mathbf{F}\mathrm{Diag}(\mathbf{y})\mathbf{u}^{(t)}\right)\right].
\end{equation}
It will be demonstrated later that this method converges faster owing to its tighter majorization.
\begin{algorithm}
	\caption{Algorithm for Phase Noise Estimation with TQM.}
	\label{alg:time_domain_PN_noprior_h_given_S}
	\begin{algorithmic}[1]
		\State Given frequency-domain transmitted symbols $\mathbf{s}$ and received symbols $\mathbf{r}$, compute the time-domain received symbols $\mathbf{y}$ by IFFT. Let $\mathbf{B} =  \mathbf{S}\check{\mathbf{F}}\left(\check{\mathbf{F}}^{H}\mathbf{S}^H\mathbf{S}\check{\mathbf{F}}\right)^{-1}\check{\mathbf{F}}^{H}\mathbf{S}^H$, where $\mathbf{S}=\mathrm{Diag}(\mathbf{s})$. Set $t=0$, and initialize $\mathbf{u}^{(0)}=e^{j\boldsymbol{\theta}_0}$.
		\Repeat
		\State $\mathbf{u}^{(t+1)} = \exp\left[j\arg\left(\mathrm{Diag}(\mathbf{y})^H\mathbf{F}^{H}\mathbf{B}\mathbf{F}\mathrm{Diag}(\mathbf{y})\mathbf{u}^{(t)}\right)\right]$
		\State $t\gets t+1$
		\Until{convergence.}
	\end{algorithmic}
\end{algorithm}

The whole procedure is summarized in Algorithm \ref{alg:time_domain_PN_noprior_h_given_S} for TQM. Since main difference from TQM lies in the update of $\mathbf{u}^{(t+1)}$, the algorithm for LQM is omitted here. Once the algorithm converges to solution $\mathbf{u}^{\star}$, the phase noise estimate can be obtained by
\begin{equation}
	\hat{\boldsymbol{\theta}} = -\arg\left(\mathbf{u}^{\star}\right),
\end{equation}
and phase noise in the received time-domain OFDM symbols is compensated via
\begin{equation}
	\hat{\mathbf{y}} = \mathbf{u}^{\star}\odot\mathbf{y}.
\end{equation}

\subsubsection{Phase Rotation Ambiguity}
\label{sssec:phase_ambiguity}
From problem \eqref{eq:TD_PN_LS_prob}, it can be seen that phase noise and channel estimates are subjected to reciprocal common phase rotations. Let $\hat{\mathbf{h}}$ and $\hat{\boldsymbol{\theta}}$ be phase noise and channel estimates, respectively. The least-squares error of the estimates $\hat{\mathbf{h}}$ and $\hat{\boldsymbol{\theta}}$ is the same as that of $e^{j\theta_\mathrm{c}}\hat{\mathbf{h}}$ and $\hat{\boldsymbol{\theta}}-\theta_{\mathrm{c}}\mathbf{1}_{N_c}$. Since $\theta_{\mathrm{c}}$ keeps unchanged among subcarriers, it acts like CFO. Many effective methods can be found for CFO correction; see, e.g., \cite{moose1994technique,lin2006joint}. Assuming CFO has been eliminated before estimating phase noise, we can thus set $\theta_{\mathrm{c}}=0$. Therefore, once Algorithm \ref{alg:time_domain_PN_noprior_h_given_S} converges to a solution $\mathbf{u}^{\star}$, phase ambiguity can be removed by the rotation: $\mathbf{u}^{\star}\gets\mathbf{u}^{\star}/u_0^{\star}$.

\subsubsection{Dimensionality Reduction}
\label{sssec:dimensionality_reduction}
Dimensionality reduction has been proposed in \cite{mathecken2016ofdm} to alleviate the computational complexity when solving an SDP of size $N_c$, the number of OFDM subcarriers. More important, as noted in \cite{zou2007compensation}, estimation problem \eqref{eq:TD_PN_LS_alt_MM_prob} and equivalent \eqref{eq:TD_PN_LS_prob} are essentially underdetermined. To obtain reasonable estimates, the number of unknowns in the problem needs to be reduced and, hence, a reduced phase noise vector is estimated.

Similar to the transformation \eqref{eq:transformation_spectral} introduced in \cite{mathecken2016ofdm} for estimating reduced spectral phase noise, we apply dimensionality reduction to our problem \eqref{eq:min_LS_err_time_domain} in the time domain. Recall $\mathbf{u} = e^{-j\boldsymbol{\theta}}$ for phase noise $\boldsymbol{\theta}$. We define
\begin{equation}
\label{eq:TD_transformation}
	\mathbf{u}_N=\mathbf{T}_N\check{\mathbf{u}}_N=\mathbf{T}_Ne^{-j\boldsymbol{\check{\theta}}_N}
\end{equation} 
as a mapping from a low-dimensional phase noise $\boldsymbol{\check{\theta}}_N\in\mathbb{R}^N$ to the original phase noise with the transformation matrix $\mathbf{T}_N\in\mathbb{R}^{N_c\times N}$ ($N<N_c$). Two instances of $\mathbf{T}_N$ are suggested in \cite{mathecken2016ofdm}---piecewise-constant transformation (PCT) and random perturbator. And it has been demonstrated that PCT, albeit simple, achieves the best performance. PCT is defined as
\begin{equation}
	\label{eq:piecewise_constant_T}
	\mathbf{T}_N = \begin{bmatrix}
		\mathbf{1}_{N_s} & \mathbf{0} & \dots & \mathbf{0} \\
		\mathbf{0} & \mathbf{1}_{N_s} & \dots & \mathbf{0} \\
		\vdots & \vdots & \ddots & \vdots \\
		\mathbf{0} & \mathbf{0} & \dots & \mathbf{1}_{N_s}
	\end{bmatrix},
\end{equation}
with $N_s=N_c/N$. In this case, the transformation matrix functions as a sample-and-hold circuit to recover the desired phase noise. Another transformation matrix is provided in \cite{zou2007compensation} based on interpolation. For simplicity, we employ PCT in this paper for the purpose of dimensionality reduction.

Introducing dimensionality reduction requires us to solve a different optimization problem. By substituting \eqref{eq:TD_transformation} into \eqref{eq:min_LS_err_time_domain}, we can obtain an estimation problem of a lower dimension. A similar procedure, however, can be followed when majorizing the new objective function and developing the MM algorithms. For TQM, the update \eqref{eq:TD_MM_1_update} is modified accordingly as
\begin{equation}
	\check{\mathbf{u}}^{(t+1)}_N = e^{j\arg\left(\mathbf{T}_N^H\mathrm{Diag}(\mathbf{y})^H\mathbf{F}^{H}\mathbf{B}\mathbf{F}\mathrm{Diag}(\mathbf{y})\mathbf{T}_N\check{\mathbf{u}}^{(t)}_N\right)}.\label{eq:TD_MM_1_update_DR}
\end{equation}
For LQM, the majorization function needs to be recomputed as the condition $\lambda\mathbf{I}_N\succeq\mathbf{T}_N^H\tilde{\mathbf{V}}\mathbf{T}_N$ involves $\mathbf{T}_N$. Notice that $\lambda_{\mathrm{max}}(\tilde{\mathbf{V}})= \|\mathbf{y}\|^2_{\infty}$, and $\lambda$ can be set to be $\|\mathbf{y}\|^2_{\infty}\lambda_{\max}(\mathbf{T}_N^H\mathbf{T}_N)$ such that the majorization inequality constraint is satisfied. As a result, the update for LQM is obtained as follows:
\begin{equation}
\label{eq:TD_MM_2_update_RD}
	\check{\mathbf{u}}^{(t+1)}_N = e^{j\arg\left(\left(\|\mathbf{y}\|^2_{\infty}\lambda_{\max}(\mathbf{T}_N^H\mathbf{T}_N)\mathbf{I}_N - \mathbf{T}_N^H\tilde{\mathbf{V}}\mathbf{T}_N\right)\check{\mathbf{u}}^{(t)}_N\right)}.
\end{equation}
For our chosen PCT, $\lambda_{\max}(\mathbf{T}_N^H\mathbf{T}_N)=N_s$ and \eqref{eq:TD_MM_2_update_RD} simplifies to
\begin{equation}
\label{eq:TD_MM_2_update_RD_2}
\check{\mathbf{u}}^{(t+1)}_N = \exp\left[j\arg\left(\left(\|\mathbf{y}\|^2_{\infty}N_s\mathbf{I}_N - \mathbf{T}_N^H\tilde{\mathbf{V}}\mathbf{T}_N\right)\check{\mathbf{u}}^{(t)}_N\right)\right].
\end{equation}
Once the optimal solution $\check{\mathbf{u}}^\star_N$ is found, estimate for the original phase noise can be obtained by
\begin{equation}
	\hat{\boldsymbol{\theta}} = -\arg\left(\mathbf{T}_N\check{\mathbf{u}}^{\star}_N\right).
\end{equation}

\begin{algorithm}
	\caption{Phase Noise Estimation with TQM and the Optimal PCT Selected by BIC.}
	\label{alg:time_domain_PN_adaptive_PCT}
	\begin{algorithmic}[1]
		\State Given frequency-domain transmitted symbols $\mathbf{s}$ and received symbols $\mathbf{r}$, compute the time-domain received symbols $\mathbf{y}$ by IFFT. Let $\mathbf{B} =  \mathbf{S}\check{\mathbf{F}}\left(\check{\mathbf{F}}^{H}\mathbf{S}^H\mathbf{S}\check{\mathbf{F}}\right)^{-1}\check{\mathbf{F}}^{H}\mathbf{S}^H$, where $\mathbf{S}=\mathrm{Diag}(\mathbf{s})$. $\mathbb{T}$ is a set of PCT matrices of different reduced length $N$.
		\For {each $\mathbf{T}_N\in \mathbb{T}$}
		\State set $t=0$ and initialize $\check{\mathbf{u}}^{(0)}_N$
		\Repeat
		\State $\check{\mathbf{u}}^{(t+1)}_N = e^{j\arg\left(\mathbf{T}_N^H\mathrm{Diag}(\mathbf{y})^H\mathbf{F}^{H}\mathbf{B}\mathbf{F}\mathrm{Diag}(\mathbf{y})\mathbf{T}_N\check{\mathbf{u}}^{(t)}_N\right)}$
		\State $t\gets t+1$
		\Until{convergence}
		\State $\mathbf{u}_N=\mathbf{T}_N\check{\mathbf{u}}^{(t+1)}_N$
		\EndFor
		\State choose $\mathbf{u}_N$ with the minimal $\mathrm{BIC}\left(\mathbf{u}_N\right)$.
	\end{algorithmic}
\end{algorithm}

The transformation matrix requires the reduced dimension $N$ to be specified in advance. Previous works have assumed a fixed PCT with given $N$, which is hardly flexible to different SNRs. Here, we prescribe a set of values of $N$ and run our algorithm for each of those values. In particular, $N$ is chosen as a factor of $N_c$ such that PCT is well-defined. To choose the optimal $N$, we employ the Bayesian Information Criterion (BIC) \cite{stoica2004model}, which has been demonstrated very effective in model order selection to avoid over-fitting. For each estimate $\mathbf{u}_N$, the corresponding BIC is defined as
\begin{equation}
\label{eq:BIC_def}
	\mathrm{BIC}\left(\mathbf{u}_N\right) = -2\ln p\left(\mathbf{y},\check{\mathbf{u}}_N\right) + N\ln N_c,
\end{equation}
where $p\left(\mathbf{y},\check{\mathbf{u}}_N\right)$ is the probability density function (PDF) of $\mathbf{y}$ given $\check{\mathbf{u}}_N$. With model \eqref{eq:OFDM_time_domain_PN} and transformation \eqref{eq:TD_transformation}, \eqref{eq:BIC_def} can be rewritten as
\begin{equation}
\label{eq:BIC_trans}
	\mathrm{BIC}\left(\mathbf{u}_N\right) = \frac{\mathcal{E}(\boldsymbol{\theta})}{\sigma^2} + N\ln N_c,
\end{equation}
where $\mathcal{E}(\boldsymbol{\theta})$ is the least-squares error \eqref{eq:LS_error_theta} of the phase noise estimate $\mathbf{u}_N$. The optimal PCT is then defined as the one that produces the minimal BIC. In doing so, improved estimates are expected, compared with the traditional methods \cite{zou2007compensation,mathecken2016ofdm}. Furthermore, the computational efficiency of LQM and TQM also guarantees an acceptable computational cost. The whole procedure is described in Algorithm \ref{alg:time_domain_PN_adaptive_PCT}.

\section{Simulation Results}
\label{sec:simulaitons}
In simulations, we consider a Rayleigh fading channel of length $L=10$, where each tap is independently distributed with exponentially decreasing power of rate $0.7$ and channel noise is circularly symmetric complex Gaussian with $\sigma=0.1$. Transmitted OFDM symbols are generated randomly (assumed known to receiver) with distribution $\mathcal{CN}(\mathbf{0},2\mathbf{I})$, the number of which is 512 or 1024. To apply dimensionality reduction, we use PCT with the reduced dimension indicated by the value of $N$ \eqref{eq:piecewise_constant_T}. Throughout this section we choose $N=32$, thus
\begin{equation}
\mathbf{T}_{N} = \begin{bmatrix}
\mathbf{1}_{32} & \mathbf{0} & \dots & \mathbf{0} \\
\mathbf{0} & \mathbf{1}_{32} & \dots & \mathbf{0} \\
\vdots & \vdots & \ddots & \vdots \\
\mathbf{0} & \mathbf{0} & \dots & \mathbf{1}_{32}
\end{bmatrix}.
\end{equation}

The following methods are considered in our simulations. PNC, as the benchmark, is the algorithm proposed in \cite{mathecken2016ofdm}, where phase noise is estimated in the frequency domain by solving an SDP \eqref{eq:min_LS_err_spectral_reduced}. Another method AltOpt refers to the alternating optimization algorithm proposed in \cite{zou2007compensation}. A modified version AltMM, based on the MM, is also compared. Specifically, we have modified AltOpt to take into account the phase noise constraint in each phase noise estimate update as opposed to the original algorithm; see Appendix for details. TQM and LQM are two MM-based algorithms we have proposed to solve the time-domain problem \eqref{eq:min_LS_err_time_domain}. From a set of prespecified PCTs with different values of $N$, opt-PCT is defined as the one that gives the minimal BIC. In particular, opt-PCT is selected from PCTs with $N\in \{2^5, 2^6, \dots, 2^{\log_2 N_c}\}$. For comparison, Ignore PHN and Exact PHN are also included, where phase noise is ignored and the exact phase noise is used, respectively, for estimating channel impulse response. Whenever necessary, an algorithm is initiated with an all-ones vector and regarded converged when the $\ell_2$-norm of difference between two consecutive iterates is no larger than $10^{-8}$. The maximum number of iterations allowed toward the convergence is 1000. All simulations were run in Matlab on a PC with a 3.20 GHz i5-4570 CPU and 8 GB RAM.

\begin{figure*}[!t]
	\centering
	\subfloat[]{\includegraphics[width=\columnwidth, trim = 18 5 36 15, clip]{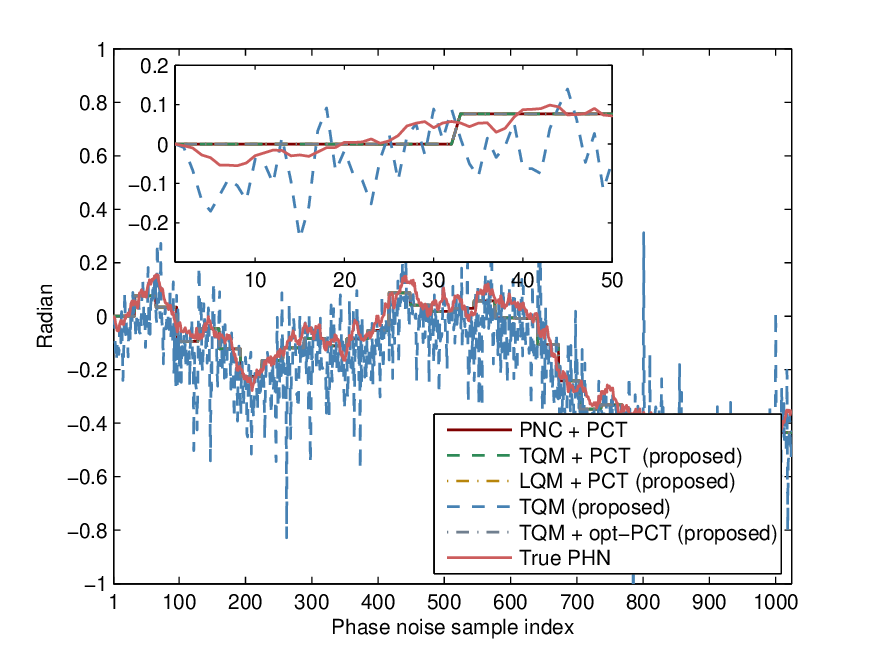}
		\label{fig:phase_noise_estimate_PN3dB_500Hz_SNR_15_Nc1024}}
	\hfill
	\subfloat[]{\includegraphics[width=\columnwidth, trim = 20 5 38 15, clip]{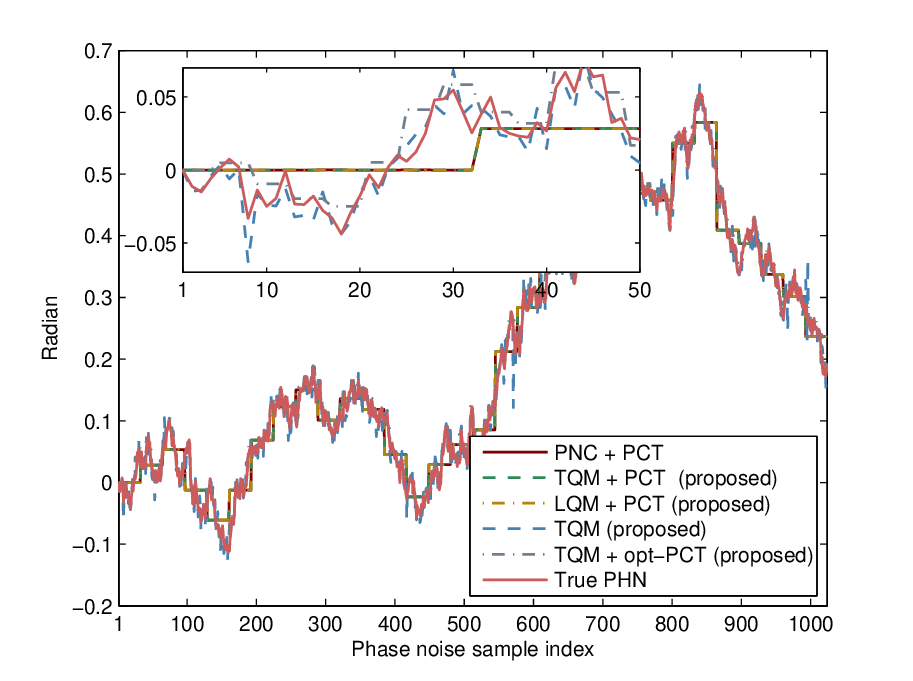}
		\label{fig:phase_noise_estimate_PN3dB_500Hz_SNR_35_Nc1024}}
	\hfill
	\subfloat[]{\includegraphics[width=\columnwidth, trim = 18 5 38 15, clip]{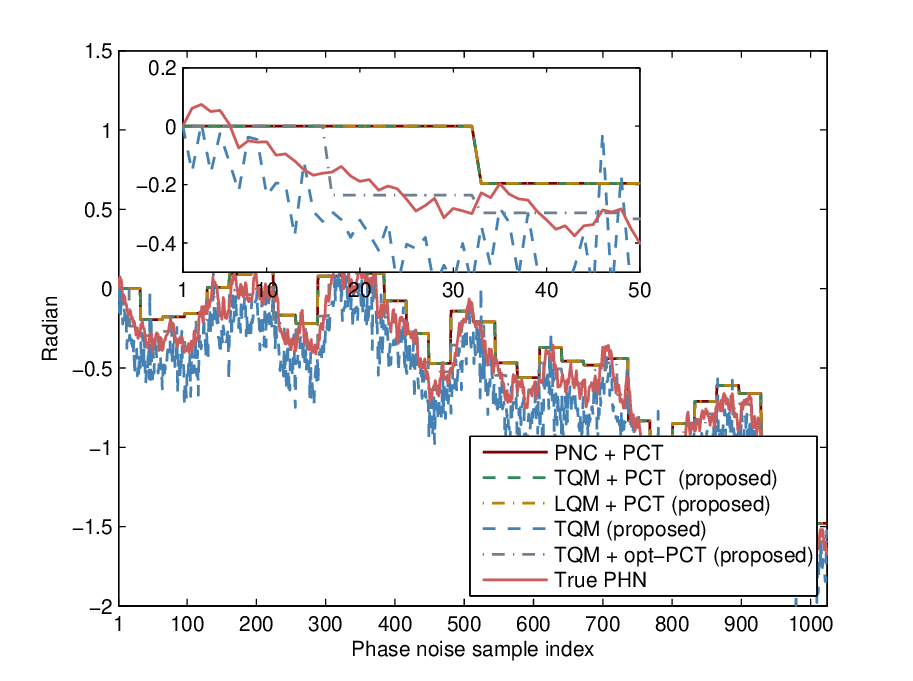}
		\label{fig:phase_noise_estimate_PN3dB_5000Hz_SNR_15_Nc1024}}
	\hfill
	\subfloat[]{\includegraphics[width=\columnwidth, trim = 20 5 38 15, clip]{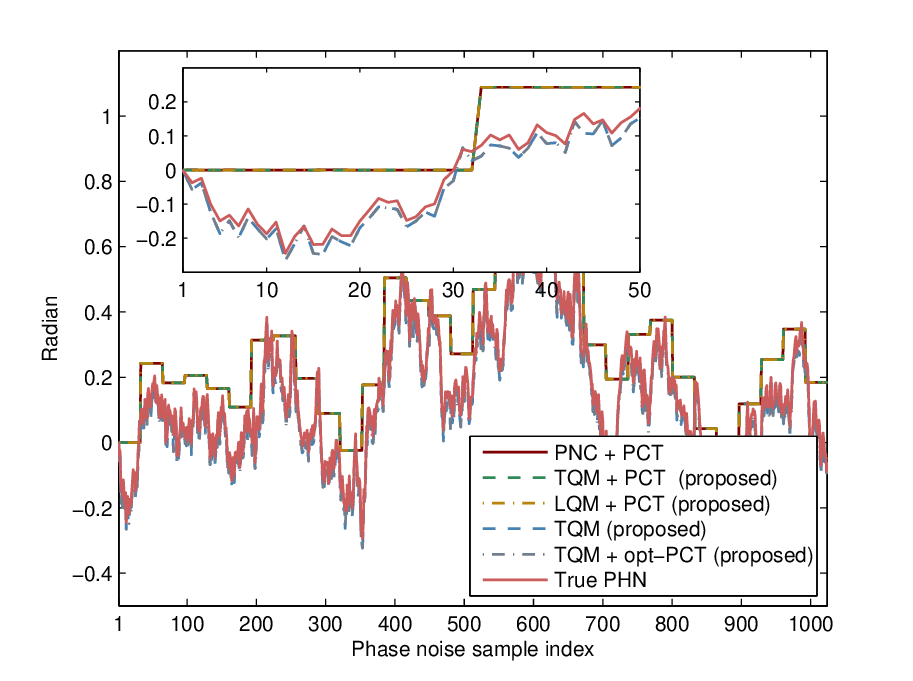}
		\label{fig:phase_noise_estimate_PN3dB_5000Hz_SNR_35_Nc1024}}
	\caption{Four instances of the estimated phase noise $\boldsymbol{\theta}$. $N_c=1024$. Where PCT is applied, $N=32$. (a) $\mathrm{SNR}=15$ dB, $\Delta f_{3\mathrm{dB}}=500$ Hz, (b) $\mathrm{SNR}=35$ dB, $\Delta f_{3\mathrm{dB}}=500$ Hz, (c) $\mathrm{SNR}=15$ dB, $\Delta f_{3\mathrm{dB}}=5000$ Hz, (d) $\mathrm{SNR}=35$ dB, $\Delta f_{3\mathrm{dB}}=5000$ Hz.\protect\footnotemark[2]}
\end{figure*}

\subsection{Phase noise and channel estimation}
In this section, we define phase noise $\boldsymbol{\theta}$ as a Wiener process \cite{mathecken2016ofdm,zou2007compensation,petrovic2007effects}. The baseband sampling rate is $f_{\mathrm{s}}=20$ MHz; 3-dB bandwidth $\Delta f_{3\mathrm{dB}}$ of phase noise is 500 Hz or 5000 Hz; assuming CFO has been fixed, i.e., $\theta_0=0$, phase noise is generated with
\begin{equation}
\theta_n-\theta_{n-1}\sim\mathcal{N}\left(0, \sqrt{\frac{2\pi\Delta f_{3\mathrm{dB}}}{f_{\mathrm{s}}}}\right),
\end{equation}
for $n=1,\dots,N_c-1$.
We first show four instances of phase noise estimates to provide an intuitive idea of how different algorithms perform, and then compare the resultant phase noise and channel estimation errors by Monte Carlo simulations.

Fig.~\subref*{fig:phase_noise_estimate_PN3dB_500Hz_SNR_15_Nc1024}--\subref*{fig:phase_noise_estimate_PN3dB_5000Hz_SNR_35_Nc1024} show phase noise estimates under four different scenarios. In all cases, PNC, TQM, and LQM yield the same estimate when the given PCT is applied. 

\subsubsection{Small Phase Noise and Low SNR}
In the small phase noise case with $\Delta f_{3\mathrm{dB}}=500$ Hz, as Fig. \subref*{fig:phase_noise_estimate_PN3dB_500Hz_SNR_15_Nc1024} shows, using the given PCT results in a staircase-like estimate; loose though it may seem, it is actually beneficial when SNR is limited, owing to the fundamental underdetermined issue of the original problem. In fact, TQM with opt-PCT provides the same estimate as that of the benchmark---opt-PCT in this case is $\mathbf{T}_{32}$. In contrast, TQM without PCT turns out an estimate with many undesired peaks associated with larger MSE. It implies, therefore, that with small phase noise and low SNR, dimensionality reduction is recommended in order to achieve a relatively better performance.
\footnotetext[2]{Since AltOpt and AltMM give almost the same MSE as PNC, we will only compare our proposed algorithms with PNC with respect to MSE. For comparison of computation time for each algorithm, see Table \ref{tab:CPU_time}.}
\begin{figure*}[!t]
	\centering
	\subfloat[]{\includegraphics[width=\columnwidth, trim = 18 5 38 15, clip]{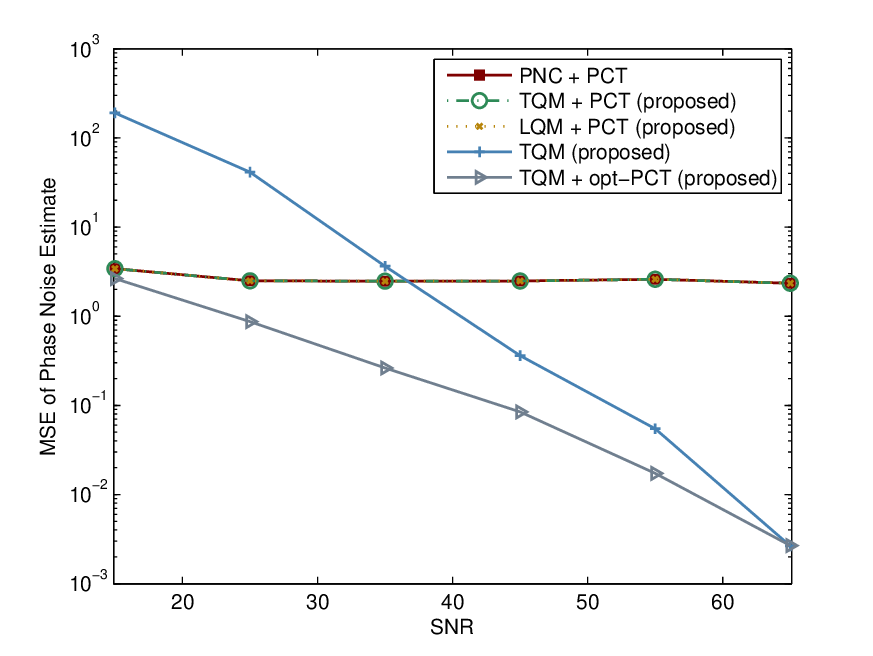}
		\label{fig:PN_MSE_PN3dB_500Hz_Nc1024_N32}}
	\hfill
	\subfloat[]{\includegraphics[width=\columnwidth, trim = 18 5 38 15, clip]{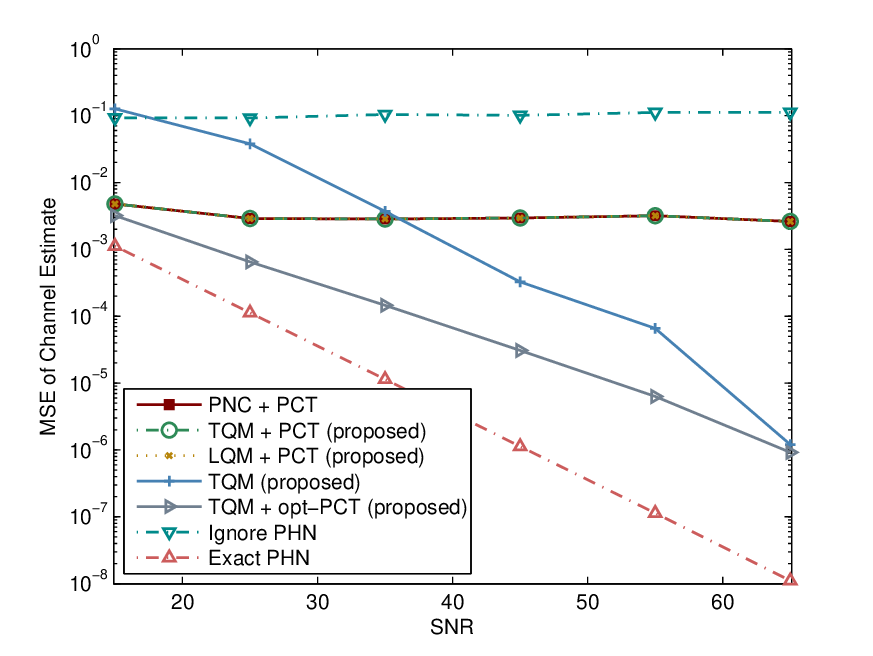}
		\label{fig:CIR_MSE_PN3dB_500Hz_Nc1024_N32}}
	\hfill
	\subfloat[]{\includegraphics[width=\columnwidth, trim = 18 5 38 15, clip]{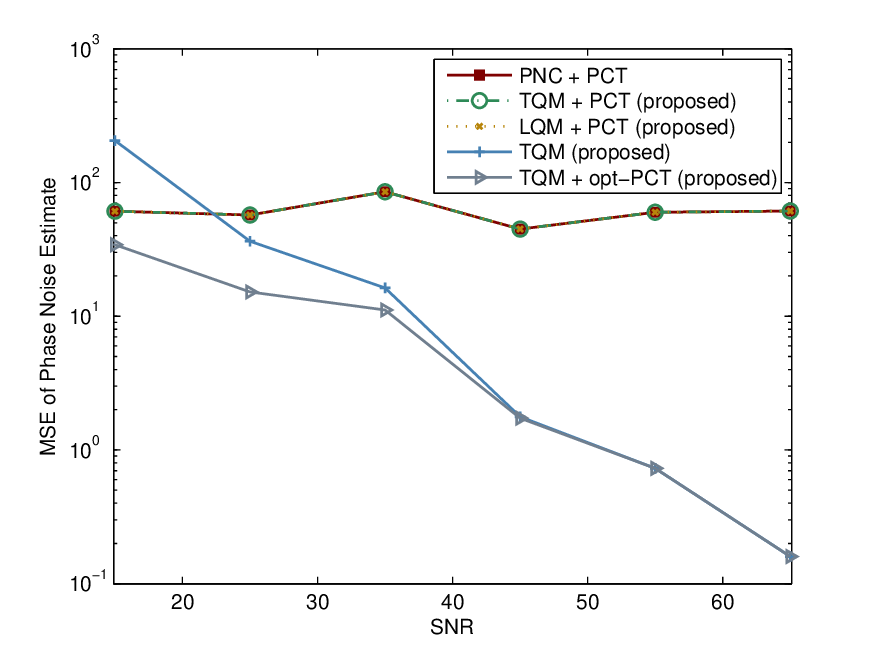}
		\label{fig:PN_MSE_PN3dB_5000Hz_Nc1024_N32}}
	\hfill
	\subfloat[]{\includegraphics[width=\columnwidth, trim = 18 5 38 15, clip]{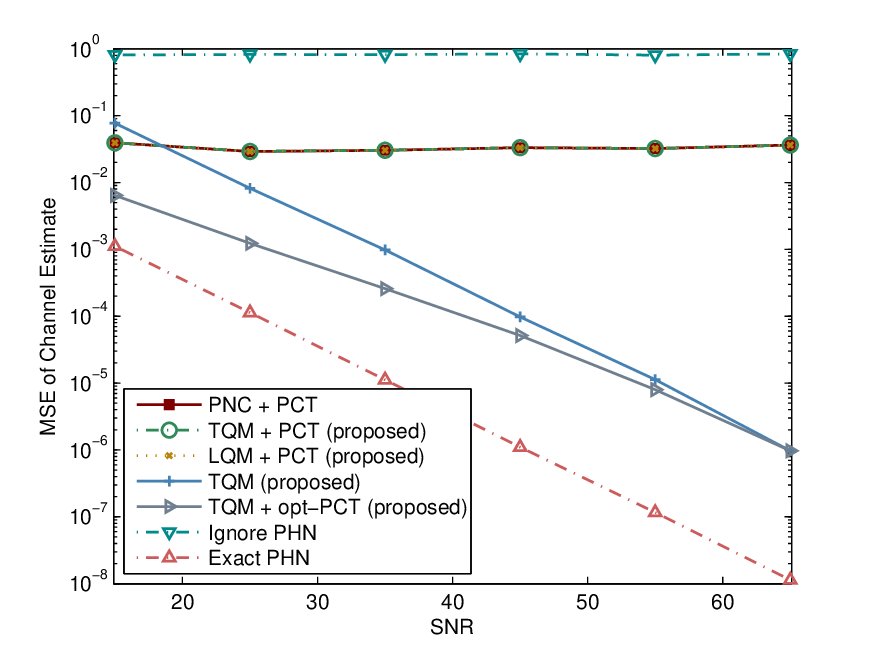}
		\label{fig:CIR_MSE_PN3dB_5000Hz_Nc1024_N32}}
	\caption{Joint estimation of phase noise and channel under different values of SNR with $N_c=1024$ and 500 Monte Carlo simulations: (a) averaged MSE of phase noise estimate, $\Delta f_{3\mathrm{dB}}=500$ Hz (b) averaged MSE of channel estimate, $\Delta f_{3\mathrm{dB}}=500$ Hz, (c) averaged MSE of phase noise estimate, $\Delta f_{3\mathrm{dB}}=5000$ Hz, (d) averaged MSE of channel estimate, $\Delta f_{3\mathrm{dB}}=5000$ Hz.}
\end{figure*}

\subsubsection{Small Phase Noise and High SNR}
A particular example for this case is shown in Fig. \subref*{fig:phase_noise_estimate_PN3dB_500Hz_SNR_35_Nc1024}. With the given PCT, we can see that PNC, TQM, and LQM are still able to produce the same good estimate by and large, the resulting MSE of which is 0.8349. TQM without PCT and TQM with opt-PCT (opt-PNC is $\mathbf{T}_{512}$ here) outperform their opponents though, with corresponding MSE being 0.1550 and 0.1890, respectively, which implies that dimension should not be reduced too much with high SNR. Particularly, no dimensionality reduction is recommended as it turns out that TQM without PCT gives a lightly better estimate. By Monte Carlo simulations, however, TQM with opt-PCT gives the minimal MSE among all the methods; see Fig. \subref*{fig:PN_MSE_PN3dB_500Hz_Nc1024_N32}. It should also be mentioned that this will pose a challenge to the benchmark PNC because it needs to deal with an SDP with size of 512 or even larger. More details will be illustrated in Section \ref{ssec:Algorithm_convergence}.

\subsubsection{Large Phase Noise and Low SNR}
A major difference in this scenario from the previous one is that TQM without PCT yields a phase noise estimate with some undesired peaks; see Fig. \subref*{fig:phase_noise_estimate_PN3dB_5000Hz_SNR_15_Nc1024}. The reason is that low SNR renders the original problem more susceptible to the underdetermined issue---the same as the case of small phase noise and low SNR in Fig. \subref*{fig:phase_noise_estimate_PN3dB_500Hz_SNR_15_Nc1024}. Nonetheless, TQM with opt-PCT gives a very good estimate outperforming other opponents; the resulting MSE for the five methods in order are 22.7332, 22.7333, 22.7374, 37.2954, and 4.7228. And opt-PCT in this example is $\mathbf{T}_{64}$.
\begin{table*}[!t]
	\centering
	\begin{threeparttable}
		\caption{CPU time of different algorithms with $\Delta f_{3\mathrm{dB}}=5000$ Hz, $\mathrm{SNR}=35$ dB, and 500 Monte Carlo simulations\tnote{\dag}}
		\label{tab:CPU_time}
		\begin{tabular}{ l c c c c c c}
			\hline
			\toprule
			& \multicolumn{6}{c}{CPU Time (s)} \\
			\cmidrule(r){2-7}
			& \multicolumn{3}{c}{$N_c=512$} & \multicolumn{3}{c}{$N_c=1024$}\\
			\cmidrule(r){2-4} \cmidrule(r){5-7}
			Algorithms                              & Mean            & Min    & Max    & Mean                & Min    & Max    \\
			\midrule
			PNC + PCT (\emph{benchmark})            & 0.5010          & 0.4459 & 0.6427 & 0.9725              & 0.8807 & 1.2901 \\
			AltOpt + PCT (\emph{benchmark})         & 0.6758          & 0.5252 & 1.0655 & 3.2983              & 2.6662 & 5.0401 \\
			AltMM + PCT (\emph{modified benchmark}) & 0.2438          & 0.2009 & 0.4077 & 0.9443              & 0.7052 & 1.3784 \\
			TQM (\emph{proposed})                   & \textbf{0.0626} & 0.0469 & 0.0973 & \textbf{0.2757}     & 0.2224 & 0.4149 \\
			LQM (\emph{proposed})                   & 0.4363          & 0.0533 & 0.8231 & 2.7519\tnote{\ddag} & 0.3243 & 5.2693 \\
			TQM + PCT (\emph{proposed})             & 0.0171          & 0.0157 & 0.0264 & 0.0658              & 0.0635 & 0.1212 \\
			LQM + PCT (\emph{proposed})             & 0.0204          & 0.0189 & 0.0284 & 0.0795              & 0.0770 & 0.1094 \\
			TQM + opt-PCT (\emph{proposed})         & \textbf{0.1367} & 0.1256 & 0.1775 & \textbf{0.7372}     & 0.6944 & 1.0055 \\
			\bottomrule
		\end{tabular}
		\begin{tablenotes}
			\item[\dag] CPU time measured for each algorithm includes computing phase noise estimate, channel estimate, and resulting MSE. For our proposed algorithms, fast Fourier transforms (FFT) is employed, wherever it is possible, to improve computational efficiency.
			\item[\ddag] LQM costs more CPU time when $N_c$ is large. But some efficient acceleration schemes can be used to boost the convergence rate, e.g., the SQUAREM method \cite{Varadhan2008}.
		\end{tablenotes}
	\end{threeparttable}
\end{table*}

\subsubsection{Large Phase Noise and High SNR}
In this last example, all the five algorithms yield nearly good estimates as Fig. \subref*{fig:phase_noise_estimate_PN3dB_5000Hz_SNR_35_Nc1024} shows. Fixed PCT for PNC, TQM, and LQM, however, still provide a relatively loose result that could have been improved with available high SNR, the MSE of which are 23.9709, 23.9709, and 23.9668, respectively. We can also expect that TQM without PCT and TQM with opt-PCT perform better, the resulting MSE of both is 0.6284. Opt-PCT in this example is $\mathbf{T}_{1024}$, which is simply an identity matrix, i.e., no dimensionality reduction is applied.

Fig. \subref*{fig:PN_MSE_PN3dB_500Hz_Nc1024_N32}--\subref*{fig:CIR_MSE_PN3dB_5000Hz_Nc1024_N32} show averaged MSE of phase noise and channel estimates with 500 Monte Carlo trials. From Fig. \subref*{fig:PN_MSE_PN3dB_500Hz_Nc1024_N32} and \subref*{fig:PN_MSE_PN3dB_5000Hz_Nc1024_N32}, we see that PNC, TQM, and LQM are comparable to each other with dimensionality reduction when SNR is low. In contrast, TQM without PCT can provide much better phase noise estimates for high enough SNR. A similar result can be found for channel estimation, where the resulting MSE has also been significantly improved with TQM; see Fig. \subref*{fig:CIR_MSE_PN3dB_500Hz_Nc1024_N32} and \subref*{fig:CIR_MSE_PN3dB_5000Hz_Nc1024_N32}. Generally speaking, TQM produces better estimates with opt-PCT than without PCT. And a closer look will reveal that as SNR grows larger, performance gap between these two methods shrinks. And even though the benchmark PNC can deal with larger SNR, the computational burden will be prohibitive, not to say the additional computational issues; see the following remark and an illustration of CPU time consumed by each algorithm in Section \ref{ssec:Algorithm_convergence}.
 
\emph{Remark:} PNC is proposed in \cite{mathecken2016ofdm}, where the original problem is reformulated as an SDP by using the S-procedure. The authors only prove the equivalence (strong duality) between the reduced problem and its SDP reformulation. For the original problem, however, the strong duality has not been established. From simulations, the resultant estimate of frequency-domain phase noise vector is not a reasonably good solution that satisfies the spectral constraint. Also, solving an SDP only gives an intermediate solution that requires an additional eigendecomposition step. PNC can thus easily fall within infeasibility and singularity issues when dimension is not reduced enough. These issues, however, have not been addressed in \cite{mathecken2016ofdm}. Furthermore, they have not given an explicit comparison in their paper about the MSE of phase noise or channel estimates---although they are optimizing such objectives---but show the bit-error-rate (BER) after channel encoding/decoding. Also, what they have simulated involves data detection, whereupon iterations are performed among phase noise estimation, channel estimation, and data detection.

\begin{figure}[!t]
	\centering
	\includegraphics[width=\columnwidth,trim = 16 5 38 15]{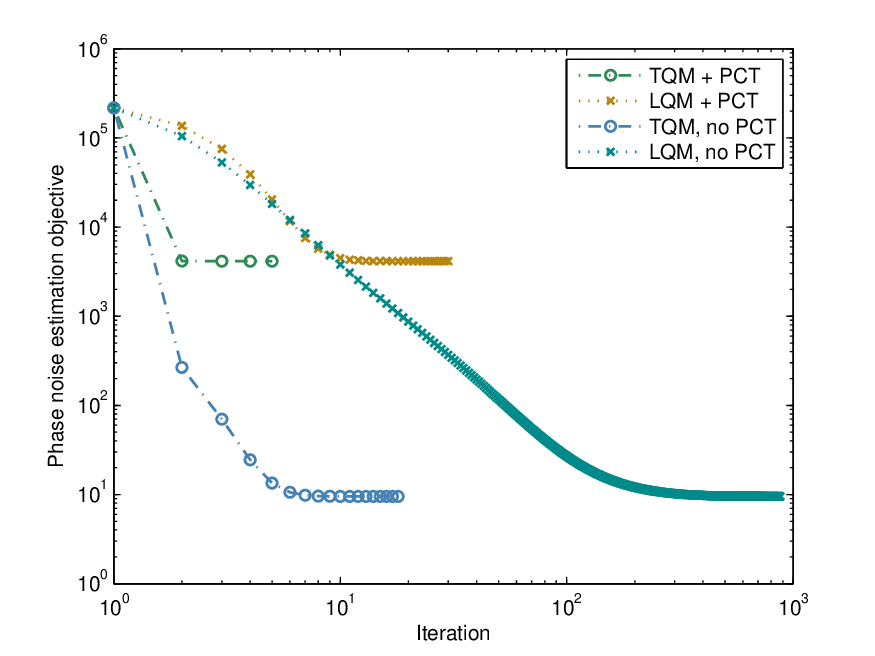}
	\caption{Convergence of TQM and LQM with and without PCT. $N_c=1024$. $\Delta f_{3\mathrm{dB}}=5000$ Hz.}
	\label{fig:cvg_TD_MM}
\end{figure}

\subsection{Algorithm convergence}
\label{ssec:Algorithm_convergence}
In this section, we first present an example of convergence properties of our proposed algorithms, and then give a comparison of CPU time consumed by each algorithm. Convergence criteria defined previously apply here as well.

Fig. \ref{fig:cvg_TD_MM} demonstrates convergence of four methods. TQM and LQM converge to the same optimal solution with the same initialization no matter PCT is applied or not. TQM, however, converges remarkably much faster than LQM, within twenty iterations or fewer. This is because TQM employs a much tighter majorization function to the original objective function. In consequence, we adopt TQM with opt-PCT in previous simulations to achieve the same performance with respect to estimation error and to save much computation time at the same time. On the other hand, as shown in previous examples, applying PCT in the case of large phase noise and high SNR causes loss of quality in the obtained estimates, which is substantiated by the fact that without PCT, much lower objective value can be achieved.
\begin{figure}[!t]
	\centering
	\subfloat[]{\includegraphics[width=\columnwidth, trim = 18 5 38 15, clip]{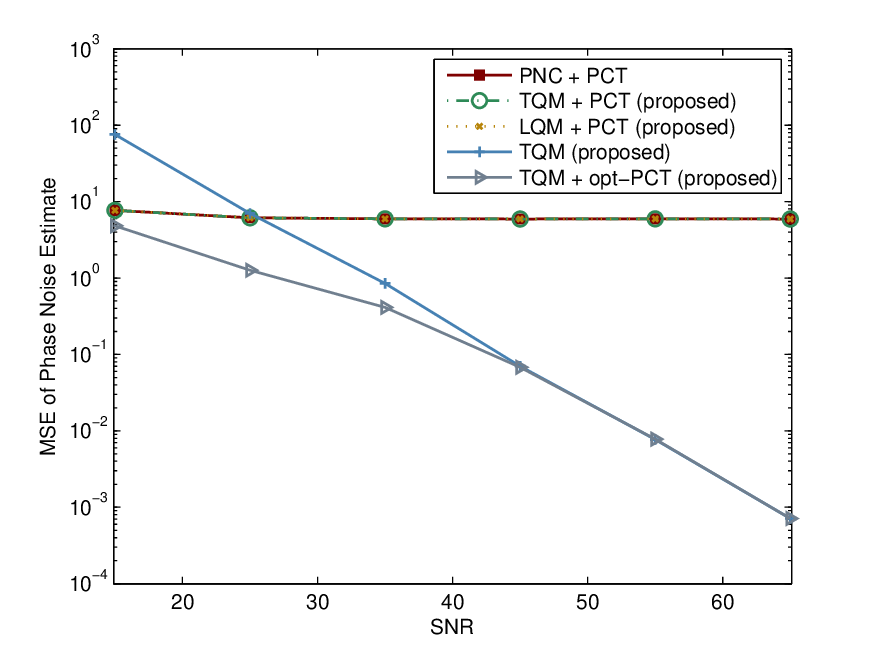}
		\label{fig:PN_MSE_PN3dB_5000Hz_Nc512_N32}}
	\hfill
	\subfloat[]{\includegraphics[width=\columnwidth, trim = 18 5 38 15, clip]{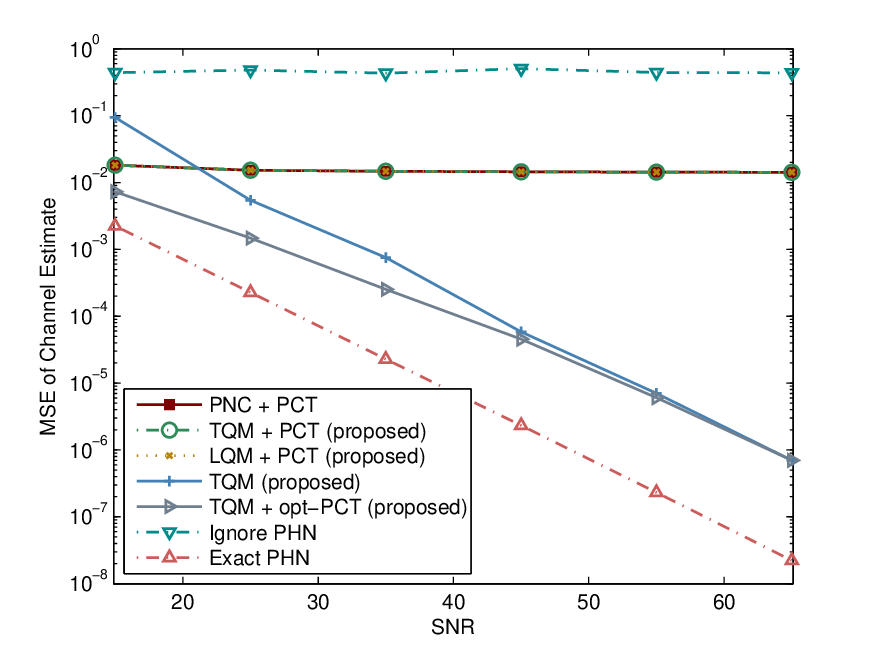}
		\label{fig:CIR_MSE_PN3dB_5000Hz_Nc512_N32}}
	\caption{Joint estimation of phase noise and channel with different values of SNR with $N_c=512$, $\Delta f_{3\mathrm{dB}}=5000\ \mathrm{Hz}$, and 500 Monte Carlo simulations: (a) averaged MSE of phase noise estimate, (b) averaged MSE of channel estimate.}
\end{figure}

A further comparison in terms of the computational complexity between our proposed algorithms and the benchmark methods is provided in Table \ref{tab:CPU_time}. Except LQM, our proposed algorithms consume less time than the benchmarks. In the case with PCT applied, our proposed algorithms outperform PNC and AltMM by saving much time and at the same time achieve the same MSE of phase noise and channel estimates as shown in the previous examples. TQM gives as the same estimate as LQM; however, it is much more efficient owing to the tighter majorization function. Despite extra time cost without PCT, TQM with opt-PCT offer the best estimate among all the methods; still, it consumes much less time than the benchmarks. In fact, when SNR is high enough, e.g., $\mathrm{SNR}=35$ dB here, TQM without PCT can still significantly improve the resulting MSE, which makes it an acceptable candidate as a suboptimal method.

\begin{figure}[!t]
	\centering
	\includegraphics[width=\columnwidth, trim = 18 5 38 15, clip]{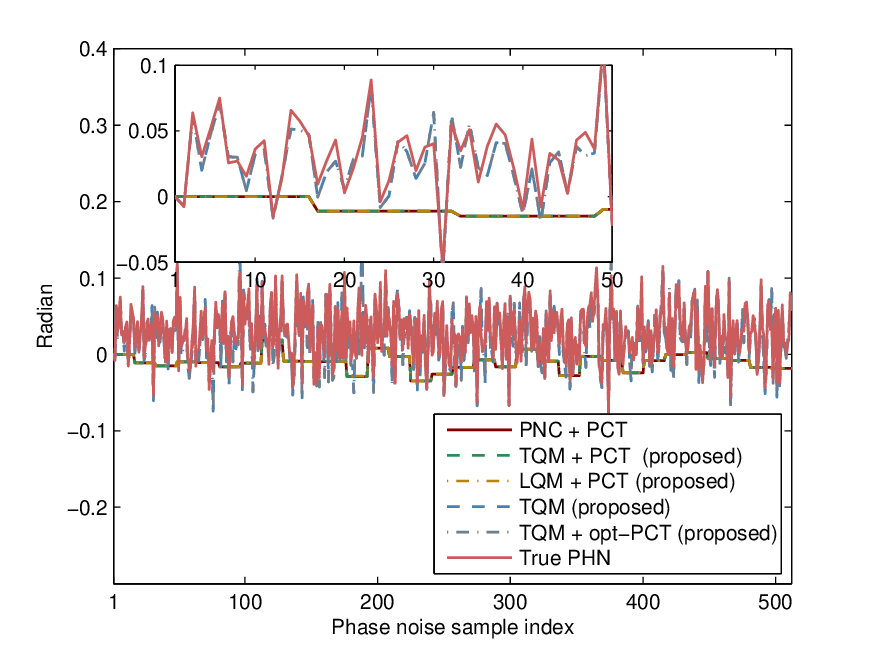}
	\caption{Gaussian phase noise estimation with $N_c=512$ and $\mathrm{SNR}=35\ \mathrm{dB}$. Where PCT is applied, $N=32$.}
	\label{fig:Gaussian_phase_noise_PN3dB_500Hz_SNR_35_Nc512}
\end{figure}
\begin{figure}[!t]
	\centering
	\subfloat[]{\includegraphics[width=\columnwidth, trim = 18 5 38 15, clip]{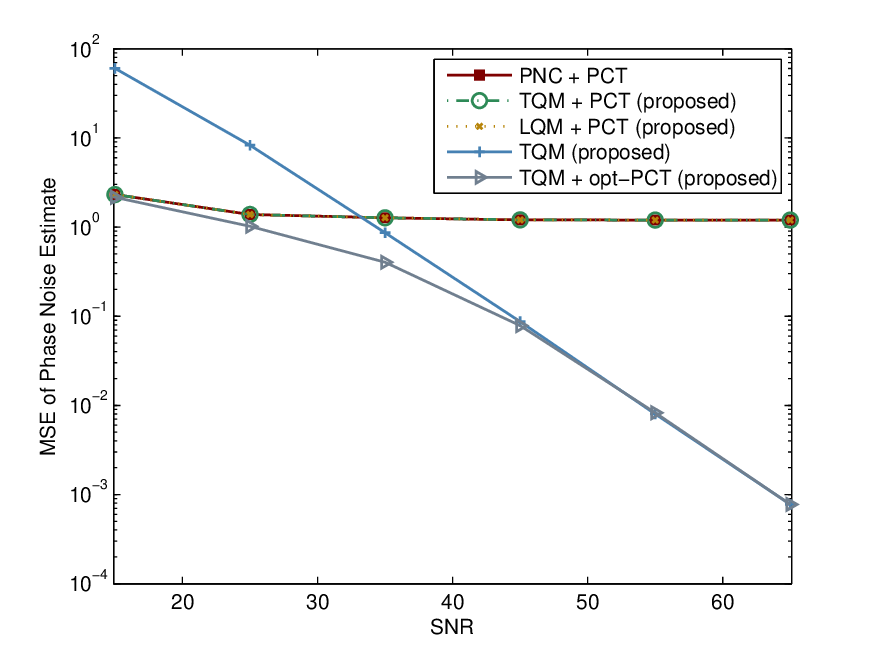}
		\label{fig:GauPN_PN_MSE_Nc512}}
	\hfill
	\subfloat[]{\includegraphics[width=\columnwidth, trim = 18 5 38 15, clip]{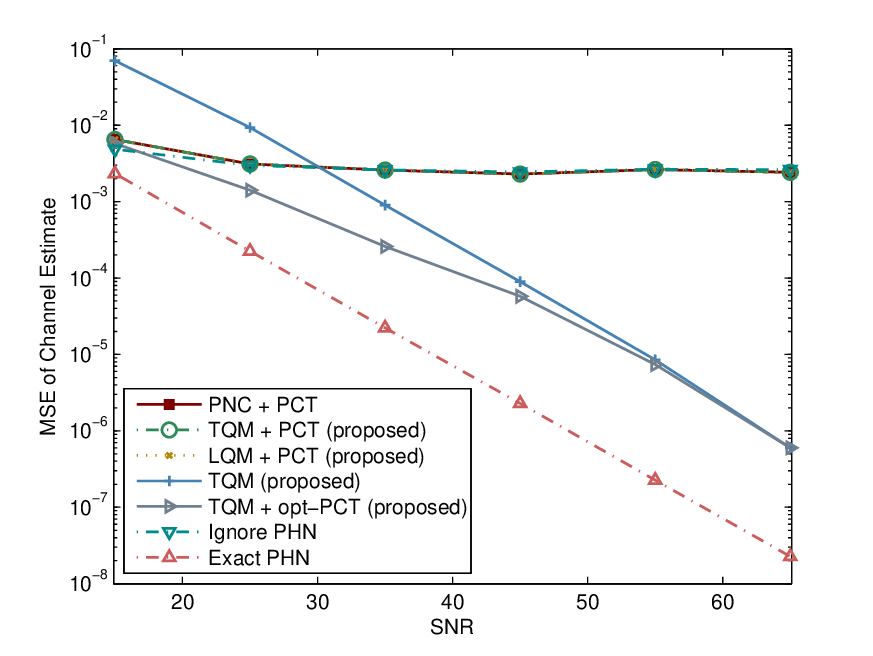}
		\label{fig:GauPN_CIR_MSE_Nc512}}
	\caption{Joint estimation of Gaussian phase noise and channel with different values of SNR with $N_c=512$ and 500 Monte Carlo simulations: (a) averaged MSE of phase noise estimate, (b) averaged MSE of channel estimate.}
\end{figure}

\subsection{Two More Examples}
\label{ssec:A_Few_More_Examples}
We present two more examples featuring fewer subcarriers deployment and the other phase noise model suggested in the literature.

\subsubsection{Wiener Phase Noise with Fewer Subcarriers}
\label{sssec:Wiener_Fewer}
Following previous examples of estimating phase noise that is modeled as a Wiener process, we show how those algorithms work with fewer subcarriers deployed. Fig. \subref*{fig:PN_MSE_PN3dB_5000Hz_Nc512_N32} displays the averaged MSE of phase noise estimates with $N_c=512$. Expectedly, the result is similar to that of Fig. \subref*{fig:PN_MSE_PN3dB_500Hz_Nc1024_N32} and \subref*{fig:PN_MSE_PN3dB_5000Hz_Nc1024_N32} in spite of either the number of subcarriers or magnitude of phase noise. Still, TQM with opt-PCT yields much better channel estimates than the benchmark, which corroborates the effectiveness of our proposed algorithms.

\subsubsection{Gaussian Phase Noise Estimation}
\label{sssec:gaussian_PN}
Phase noise generated in a phase-locked loop is modeled as a Gaussian process \cite{lin2006joint}. In this example, the number of subcarriers is $N_c=512$ with baseband  sampling rate $f_\mathrm{s}=20$ MHz. The standard deviation $\theta_\mathrm{rms}$ of phase noise generated by a phase-locked voltage controlled oscillator is 2 degrees. The single-pole butterworth filter with 3-dB bandwidth $\Delta f_{3\mathrm{dB}}=100$ Hz is adopted so that the covariance matrix of phase noise is
\begin{equation}
	\mathbf{C}_{i,j}=\left(\frac{\pi \theta_{\mathrm{rms}}}{180}\right)^2e^{\frac{-2\pi \Delta f_{3\mathrm{dB}}|i-j|}{f_\mathrm{s}}}.
\end{equation}
An instance of Gaussian phase noise and its estimates is shown in Fig. \ref{fig:Gaussian_phase_noise_PN3dB_500Hz_SNR_35_Nc512}. As its name indicates, Gaussian phase noise will not drift away too much like Wiener phase noise. Similar to Fig. \subref*{fig:phase_noise_estimate_PN3dB_500Hz_SNR_35_Nc1024} and \subref*{fig:phase_noise_estimate_PN3dB_5000Hz_SNR_35_Nc1024}, large dimensionality reduction induces considerable loss in the obtained estimates; with PCT, MSE for PNC, TQM, and LQM are 1.4049, 1.4049, and 1.4024, respectively. TQM without PCT and TQM with opt-PCT achieve the best performance in this example, both of which have the same MSE of 0.1556. And opt-PCT is just an identity matrix, which causes no dimension to be reduced. Fig. \subref*{fig:GauPN_PN_MSE_Nc512} and \subref*{fig:GauPN_CIR_MSE_Nc512} show MSE of the obtained estimates of Gaussian phase noise and channel, respectively. Still, our proposed algorithms can provide much better estimates as in the previous examples with Wiener phase noise.

\section{Conclusion}
\label{sec:Conclusion}
We have proposed two efficient algorithms and a method with dimensionality reduction coupled with the Bayesian Information Criterion for the joint phase noise and channel estimation in OFDM. The algorithms are devised based on the majorization-minimization technique and applied to two canonical models of phase noise---Wiener process and Gaussian process. The simulation results have shown that when the same dimensionality reduction is employed, our proposed algorithms achieve the same MSE as that of the benchmark but consume much less time. By further selecting the optimal dimensionality reduction with BIC, our proposed algorithms provide significantly better estimates when SNR is at least moderate but still demand no much additional computation time. Therefore, the advantage of our methods should  be outstanding in modern applications of OFDM, where a large number of subcarriers are deployed.


\appendix
\section*{Alternating Optimization with the MM}
\label{sec:Alternating_Optimization_with_MM}
In the following, the constraint of phase noise is taken into account and the alternating optimization scheme is correspondingly modified.

With the unimodular constraint for $\mathbf{c} = e^{j\mathbf{\boldsymbol{\theta}}}$, we have $\big(\boldsymbol{\Phi}_{\boldsymbol{\phi}}^{(i-1)}\big)^H\boldsymbol{\Phi}_{\boldsymbol{\phi}}^{(i-1)}=N_c\mathbf{I}$ \eqref{eq:Freq_property_PN_matrix}. And the channel estimate is updated by
\begin{equation}
\label{eq:h_update_appendix}
\hat{\mathbf{h}}^{(i)} = 		\left(N_c\check{\mathbf{F}}^{H}\mathbf{S}^{H}\mathbf{S}\check{\mathbf{F}}\right)^{-1}\check{\mathbf{F}}^{H}\mathbf{S}^{H}\left(\boldsymbol{\Phi}_{\boldsymbol{\phi}}^{(i-1)}\right)^{H}\mathbf{r}.
\end{equation}
Substitute \eqref{eq:h_update_appendix} into the objective of problem \eqref{eq:TD_PN_LS_alt_MM_prob}, and the following problem is obtained:
\begin{equation}
\label{eq:LS_alternating_PN_update_prob}
\begin{array}{rl}
\underset{\mathbf{c}:|c|_n=1, n=1,\dots,N_c}{\text{minimize}} & \left\|\mathbf{r}-\mathbf{P}\mathbf{F}\mathbf{c}\right\|^2,
\end{array}
\end{equation}
where $\mathbf{P}=\mathrm{circ}(\mathbf{S}\check{\mathbf{F}}\hat{\mathbf{h}}^{(i)})$. Instead of updating $\mathbf{c}$ by the least-squares solution \eqref{eq:least_squares_c}, the MM method can be used to solve problem \eqref{eq:LS_alternating_PN_update_prob}. The majorization can be obtained as follows \cite[Lemma 1]{song2014optimal}:
\begin{align}
\|\mathbf{r}-\mathbf{P}\mathbf{F}\mathbf{c}\|^2
=&\;
\mathbf{r}^H\mathbf{r} - 2\mathrm{Re}\big\{\mathbf{r}^H\mathbf{P}\mathbf{F}\mathbf{c}\big\} + \mathbf{c}^H\mathbf{F}^H\mathbf{P}^H\mathbf{P}\mathbf{F}\mathbf{c}\\
\leq&\;\mathbf{r}^H\mathbf{r} - 2\mathrm{Re}\big\{\mathbf{r}^H\mathbf{P}\mathbf{F}\mathbf{c}\big\} + \lambda\mathbf{c}^H\mathbf{c}\nonumber\\
& + 2\mathrm{Re}\Big\{\big(\mathbf{c}^{(t)}\big)^H\left(\mathbf{F}^H\mathbf{P}^H\mathbf{P}\mathbf{F}-\lambda\mathbf{I}\right)\mathbf{c}\Big\}\nonumber\\
& + \big(\mathbf{c}^{(t)}\big)^H\left(\lambda\mathbf{I}-\mathbf{F}^H\mathbf{P}^H\mathbf{P}\mathbf{F}\right)\mathbf{c}^{(t)}\label{eq:LS_alternating_PN_update_prob_MMupdate}.	
\end{align}
To obtain a good majorization function, we can choose $\lambda$ as 
\begin{align}
\lambda & = \lambda_{\max}\left(\mathbf{F}^H\mathbf{P}^H\mathbf{P}\mathbf{F}\right)\\
& = \lambda_{\max}\left(\mathbf{P}^H\mathbf{P}\right)\\
& = N_c\left\|\mathbf{F}^H\left(\mathbf{S}\check{\mathbf{F}}\hat{\mathbf{h}}^{(i)}\right)\right\|_\infty^2,
\end{align}
where Lemma \ref{lemma2} is applied to compute the maximum eigenvalue of $\mathbf{P}$. Minimizing \eqref{eq:LS_alternating_PN_update_prob_MMupdate} results in the update of $\mathbf{c}$:
\begin{equation}
\mathbf{c}^{(t+1)} = e^{j\arg \mathbf{a}^{(t+1)}},
\end{equation}
where $\mathbf{a}^{(t+1)} = \mathbf{F}^H\mathbf{P}^H\mathbf{r}+\lambda \mathbf{c}^{(t)} -  \mathbf{F}^H\mathbf{P}^H\mathbf{P}\mathbf{F}\mathbf{c}^{(t)}$. The complete procedure is described in Algorithm \ref{alg:LS_alternating_MM}. Algorithm \ref{alg:LS_alternating_MM} can also be readily modified to incorporate PCT, for which a similar majorization approach can be followed.
\begin{algorithm}[!t]
	\caption{Phase Noise Estimation by Alternating Minimization and the MM.}
	\label{alg:LS_alternating_MM}
	\begin{algorithmic}[1]
		\State Given frequency-domain transmitted symbols $\mathbf{s}$ and received symbols $\mathbf{r}$, and  $\mathbf{S}=\mathrm{Diag}(\mathbf{s})$, set $i=1$, and initialize $\hat{\mathbf{c}}^{(0)}=e^{j\boldsymbol{\theta}_0}$.
		\Repeat
		\State $\boldsymbol{\Phi}_{\boldsymbol{\phi}}^{(i-1)} = \mathrm{circ}(\mathbf{F}\hat{\mathbf{c}}^{(i-1)})$
		\State $\hat{\mathbf{h}}^{(i)} = 		\left(N_c\check{\mathbf{F}}^{H}\mathbf{S}^{H}\mathbf{S}\check{\mathbf{F}}\right)^{-1}\check{\mathbf{F}}^{H}\mathbf{S}^{H}\big(\boldsymbol{\Phi}_{\boldsymbol{\phi}}^{(i-1)}\big)^{H}\mathbf{r}$
		\State $\mathbf{P}=\mathrm{circ}\big(\mathbf{S}\check{\mathbf{F}}\hat{\mathbf{h}}^{(i)}\big)$
		\State $\lambda=N_c\big\|\mathbf{F}^H(\mathbf{S}\check{\mathbf{F}}\hat{\mathbf{h}}^{(i)})\big\|_\infty^2$
		\State $t=0$, and initialize $\mathbf{c}^{(0)}$
		\Repeat
		\State $\mathbf{a}^{(t+1)} = \mathbf{F}^H\mathbf{P}^H\mathbf{r}+\lambda \mathbf{c}^{(t)} -  \mathbf{F}^H\mathbf{P}^H\mathbf{P}\mathbf{F}\mathbf{c}^{(t)}$
		\State $\mathbf{c}^{(t+1)} = e^{j\arg \mathbf{a}^{(t+1)}}$
		\Until{convergence}
		\State $\hat{\mathbf{c}}^{(i)}=\mathbf{c}^{(t+1)}$
		\State $t\gets t+1$
		\Until{convergence.}
	\end{algorithmic}
\end{algorithm}


\bibliographystyle{IEEEtran}

\end{document}